\documentclass[preprint,12pt]{elsarticle}

\usepackage{amsmath}

\usepackage[]{subcaption}
\usepackage{hyperref}
\usepackage{cleveref}
\usepackage{xspace}
\usepackage{booktabs}
\usepackage{enumitem}

\usepackage[disable]{todonotes}

\newcommand{\ttop}{\textsc{Top}\xspace}
\newcommand{\bbottom}{\textsc{Bottom}\xspace}
\newcommand{\lleft}{\textsc{Left}\xspace}
\newcommand{\rright}{\textsc{Right}\xspace}
\newcommand{\ffree}{\textsc{Free}\xspace}

\newcommand{\bfs}{\textsc{BFS}\xspace}
\newcommand{\fd}{\textsc{FD}\xspace}
\newcommand{\rand}{\textsc{Rand}\xspace}

\newcommand{\vertices}{\textsc{Vertices}\xspace}
\newcommand{\ports}{\textsc{Ports}\xspace}
\newcommand{\mixed}{\textsc{Mixed}\xspace}
\newcommand{\kieler}{\textsc{Kieler}\xspace}

\newcommand{\pseudobc}{\textsc{PseudoBC}\xspace}
\newcommand{\oppositebc}{\textsc{OppositeBC}\xspace}
\newcommand{\relpos}{\textsc{RelPos}\xspace}

\newcommand{\ncr}{$n_\textrm{cr}$\xspace}
\newcommand{\nbp}{$n_\textrm{bp}$\xspace}

\newcommand{\xspan}{\operatorname{span}}
\newcommand{\xline}{\operatorname{line}}

\usepackage{tikz}
\usetikzlibrary{decorations.pathmorphing,decorations.pathreplacing,decorations.shapes,fit,positioning,shapes.geometric,calc,intersections,plotmarks}
\usepackage{pgfplots}
\usepackage{pgfplotstable}
\usepackage{adjustbox}

\pgfplotsset{compat=1.5}
\definecolor{cb-Set1-1}{RGB}{228,26,28}
\definecolor{cb-Set1-2}{RGB}{55,126,184}
\definecolor{cb-Set1-3}{RGB}{77,175,74}
\definecolor{cb-Set1-4}{RGB}{152,78,163}
\definecolor{cb-Set1-5}{RGB}{255,127,0}
\definecolor{cb-Set1-6}{RGB}{255,255,51}
\definecolor{cb-Set1-7}{RGB}{166,86,40}
\definecolor{cb-Set1-8}{RGB}{247,129,191}
\definecolor{cb-Set1-9}{RGB}{153,153,153}
\definecolor{cb-Dark2-1}{RGB}{27,158,119}
\definecolor{cb-Dark2-2}{RGB}{217,95,2}
\definecolor{cb-Dark2-3}{RGB}{117,112,179}
\definecolor{cb-Dark2-4}{RGB}{231,41,138}
\definecolor{cb-Dark2-5}{RGB}{102,166,30}
\definecolor{cb-Dark2-6}{RGB}{230,171,2}
\definecolor{cb-Dark2-7}{RGB}{166,118,29}
\definecolor{cb-Dark2-8}{RGB}{102,102,102}
\begin{document}

\begin{frontmatter}

%% Title, authors and addresses

%% use the tnoteref command within \title for footnotes;
%% use the tnotetext command for theassociated footnote;
%% use the fnref command within \author or \address for footnotes;
%% use the fntext command for theassociated footnote;
%% use the corref command within \author for corresponding author footnotes;
%% use the cortext command for theassociated footnote;
%% use the ead command for the email address,
%% and the form \ead[url] for the home page:
%% \title{Title\tnoteref{label1}}
%% \tnotetext[label1]{}
%% \author{Name\corref{cor1}\fnref{label2}}
%% \ead{email address}
%% \ead[url]{home page}
%% \fntext[label2]{}
%% \cortext[cor1]{}
%% \affiliation{organization={},
%%             addressline={},
%%             city={},
%%             postcode={},
%%             state={},
%%             country={}}
%% \fntext[label3]{}

\title{Layered Drawing of Undirected Graphs \\ with Generalized Port
	Constraints}

%% use optional labels to link authors explicitly to addresses:
\affiliation[uw]{organization={Institut f{\"u}r Informatik, Universit{\"a}t W{\"u}rzburg},
             city={W{\"u}rzburg},
             country={Germany}}

\affiliation[denk]{organization={denkbares GmbH},
             city={W{\"u}rzburg},
             country={Germany}}
         
\fntext[fund]{J.Z.\ acknowledges support by BMWi (ZIM project iPRALINE~-- grant ZF4117505).}

\author[uw,fund]{Johannes Zink}
\author[uw]{Julian Walter}
\author[uw,denk]{Joachim Baumeister}
\author[uw]{Alexander Wolff}

%\affiliation{organization={},%Department and Organization
%            addressline={}, 
%            city={},
%            postcode={}, 
%            state={},
%            country={}}

\begin{abstract}
The aim of this research is a practical method to draw cable plans
of complex machines.  Such plans consist of electronic components
and cables connecting specific ports of the components.  Since the
machines are configured for each client individually, cable plans
need to be drawn automatically.  The drawings must be well readable
so that technicians can use them to debug the machines.  In order to
model plug sockets, we introduce \emph{port groups}; within a group,
ports can change their position (which we use to improve the
aesthetics of the layout), but together the ports of a group must
form a contiguous block.

We approach the problem of drawing such cable plans by extending the
well-known Sugiyama framework such that it incorporates ports and
port groups.  Since the framework assumes directed graphs,
we propose several ways to orient the edges of the given undirected
graph.  We compare these methods experimentally, both on
real-world data and synthetic data that carefully simulates
real-world data.  We measure the aesthetics of the resulting
drawings by counting bends and crossings.  Using these metrics,
we experimentally compare our approach to \emph{Kieler} [JVLC 2014], a library for
drawing graphs in the presence of port constraints.
Our method produced 10--30~\% fewer crossings,
while it performed equally well or slightly worse than Kieler
with respect to the number of bends and the time used to
compute a drawing.
\end{abstract}

\begin{keyword}
%% keywords here, in the form: keyword \sep keyword
Sugiyama framework \sep port constraints \sep experimental evaluation.

%% PACS codes here, in the form: \PACS code \sep code

%% MSC codes here, in the form: \MSC code \sep code
%% or \MSC[2008] code \sep code (2000 is the default)

\end{keyword}

\end{frontmatter}

\section{Introduction}

Today, the development of industrial machinery implies a high
interdependency of mechanical, electrical, hydraulic, and
software-based components.  The continuous improvement of these
machines yielded an increased complexity in all these domains, but
also in their interrelations.
In the case of a malfunction, a human technician needs to understand
the particular interdependencies.  Only then, (s)he will be able to
find, understand, and resolve errors.  Different types of schematics
play a key role in this diagnosis task for depicting dependencies
between the involved components, e.g., electric or functional
schematics.  The intuitive understanding and comprehensibility of
these schematics is critical for finding errors efficiently.

\begin{figure}[t]
	\centering
	\includegraphics[width=\textwidth]{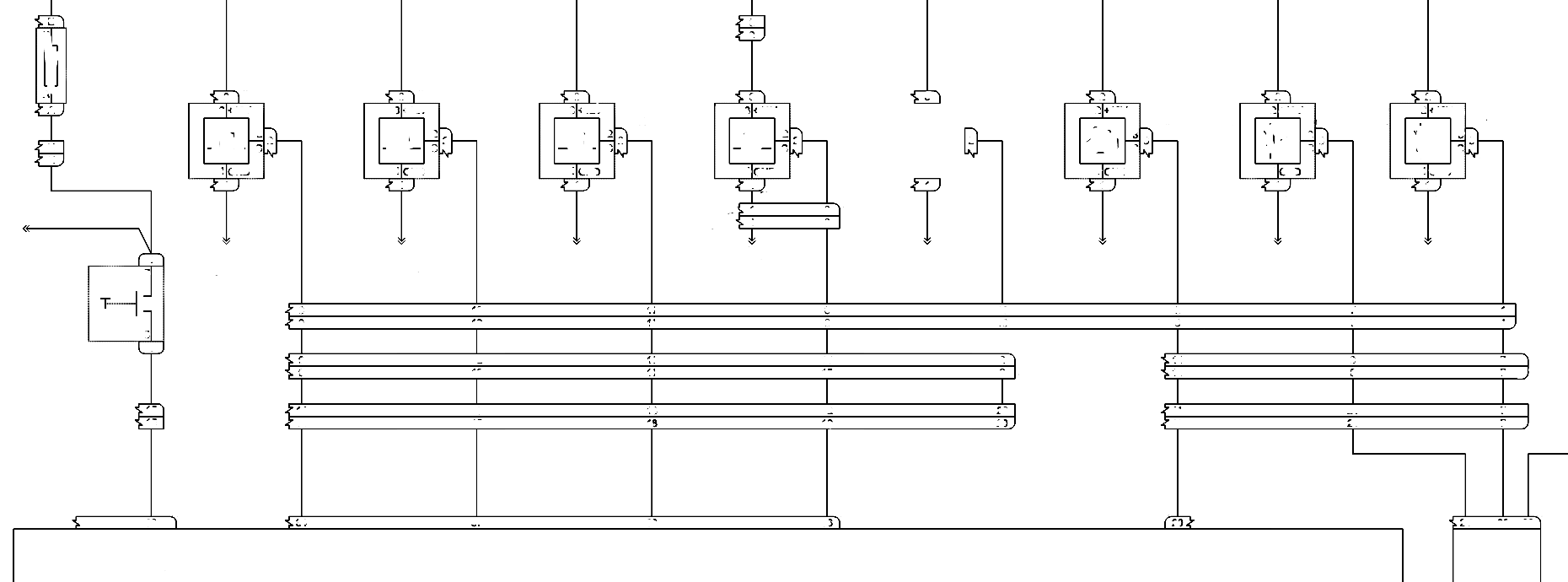}
	\caption{Extract of a hand-drawn plan.
		The labels have been intentionally obfuscated or removed.}
	\label{fig:handdrawn}
\end{figure}

Due to the increased complexity of machinery, such schematics cannot
be drawn manually anymore: The high variance of machine configurations
%(industry 4.0, lot size~1)
nowadays requires the ad-hoc computation
and visualization of schematics appropriate for the requested
diagnosis case.
To support technicians, algorithms for drawing schematics should
adhere to the visual ``laws'' of the manual drawings that the
technicians are familiar with; see Fig.~\ref{fig:handdrawn} for an example.
Such drawings route connections
between components in an orthogonal manner.
Manual drawings often use few layers and seem to avoid crossings and
bends as much as possible. \todo{Reviewer 1: give a reference}

In many applications (such as UML diagrams or data flow diagrams),
connections are directed from left to right or from top to bottom.
This setting is supported by the framework introduced by
Sugiyama et al.~\cite{Sugiyama1981}.  Given a directed graph,
their approach arranges the edges mainly in the same direction by organizing
the nodes in subsequent layers (or levels).  The layer-based approach
solves the graph-layout problem by dividing it into five phases: cycle
elimination, layer assignment, crossing minimization, node placement,
and edge routing.

There are also algorithms for practical applications purely based on
the orthogonal drawing paradigm, where all vertices are rectangles
on a regular grid and the edges are routed along the horizontal and vertical lines of the grid.
There, a classic three-phase method dates back to Biedl et al.~\cite{Biedl2000}.

In many technical drawings (such as cable plans, UML diagrams, or data
flow diagrams), components are drawn as axes-aligned rectangles,
connections between the components are drawn as axes-aligned polygonal
chains that are attached to a component using a \emph{port}, that is,
a geometric icon that is small relative to a component and whose shape
has a specific meaning for the domain expert.  Using so-called
\emph{port constraints}, a user can insist that a connection enters a
component on a specific side---a natural requirement in many
applications.

The well-established Kieler library~\cite{ssh-dlgpc-JVLC14} implements
the Sugiyama framework.  Kieler is particularly interesting for our
application as Kieler allows the user to specify several types of port
constraints; namely, on which side of a vertex rectangle should a port
be placed, and, for each side, the exact order in which the ports
should be arranged.  Alternatively, the order is variable and can be
exploited to improve the layouts in terms of crossings and bends.
Okka et al.~\cite{Okka2021} integrate these types of port constraints
to a force-directed layouting algorithm.

We have chosen to build our algorithm for undirected graphs on the
(directed) layer-based approach instead of
an (undirected) purely orthogonal one because
the typical hand-drawn plans use only few distinct layers
to place the vertices on,
the layer-based approach seems to be better investigated in practice,
and Kieler has already proven to yield by and large pleasing results in the considered domain.

\paragraph{Our Contribution}
First, we propose two methods to direct the edges
of the given undirected graph so that we can apply the
Sugiyama framework (see \Cref{sec:algorithm});
one is based on breadth-first search, the other on a
force-directed layout.  We compare the two methods experimentally with
a simple baseline method that places the nodes of the given graph
randomly and directs all edges upward (see \Cref{sub:experiments}),
both on real-world and synthetic cable plans (see
\Cref{sub:generating}).  We claim that our approach to generate
realistic test graphs is of independent interest.  We ``perturb''
real-world instances such that, statistically, they have similar
features as the original instances.

Second, we extend the set of port constraints
that the aforementioned Kieler library allows the user to specify.  In
order to model plug sockets, we introduce \emph{port groups}; within a
group, the position of the ports is either fixed or variable.  In either
case, the ports of a group must form a contiguous block.  Port groups
can be nested.  If the order of a port group is variable, our
algorithm exploits this to improve the aesthetics of the layout.

Apart from such hierarchical constraints, we also give the user the
possibility to specify pairings between ports % or port groups
that belong to opposite sides of a vertex rectangle (top and bottom).
Such a pairing constraint enforces that the two corresponding ports 
% or port groups 
are placed at the same x-coordinates on opposite sides of the vertex
rectangle.  Pairing constraints model pairs of sockets of equal width
that are plugged into each other.

After formally defining the problem % that we have sketched above
(\Cref{sec:preliminaries}), we describe our algorithm
(\Cref{sec:algorithm}).  Finally, we present our experimental
evaluation (\Cref{sec:evaluation}).

\section{Preliminaries}
\label{sec:preliminaries}

We define the problem
\textsc{Layered Graph Drawing with Generalized Port Constraints} as follows.
For an illustration refer to Fig.%s.~\ref{fig:left-right-side-ports-after} and
~\ref{fig:inserting-dummy-vertices-before}.

\medskip

\noindent
\textbf{Given:} An undirected \emph{port graph}~$G$, which is a
5-tuple $(V, P, \mathit{PG}, \mathit{PP}, E)$, where
\begin{itemize}%[leftmargin=-13.0pt]
	\item $V$ is the set of vertices---each vertex~$v$ is associated with
	two positive numbers $w(v)$ and $h(v)$; $v$~will be represented by a rectangle of
	width at least~$w(v)$ and height at least~$h(v)$
	(to ensure a given vertex label can be accommodated),
	\item $P$ is the set of ports---each port belongs either directly to a
	vertex or indirectly through a port group (or a nested sequence of
	port groups),
	\item $\mathit{PG}$ is the set of port groups---each port group belongs to a
	side (\ttop, \bbottom, \lleft, \rright, \ffree)
	of exactly one vertex and contains a
	set of ports and port groups (not contained in another port group)
	whose order is fixed or variable,
	%% Das ergibt sich aus obigem, oder?
	% there is no port group not contained in a vertex or another port group,
	\item $\mathit{PP}$ is the set of port pairings---each port pairing consists of
	two unique ports from~$P$ that belong to the same vertex (directly
	or via port groups), and
	\item $E$ is the set of edges---each edge connects two unique ports
	from~$P$ that are contained in different vertices
	(there is at most one edge per port), and
	\item the graph where all ports are contracted into their vertices is connected.
\end{itemize}

\medskip

\noindent
\textbf{Find:} A drawing of $G$ such that
\begin{itemize}%[leftmargin=-13.0pt]
	\item no drawing elements overlap each other except that edges
	may cross each other in single points,
	\item each vertex $v \in V$ is drawn as an axis-aligned
	rectangle of width at least $w(v)$ and height at least~$h(v)$ on a horizontal layer,
	\item each port $p \in P$ is drawn as a (small, fixed-size) rectangle
	attached to the boundary of its vertex rectangle (on the specified side unless set to \ffree),
	\item when walking along the boundary of a vertex, the ports of a port
	group (or subgroup) form a contiguous block; and for a port group
	with fixed order, its ports and port groups appear in that order,
	\item for each port pair $\{p, p'\} \in \mathit{PP}$, ports $p$ and $p'$ are
	drawn on the same vertical or horizontal line on opposite sides of their vertex,
	\item each edge $\{p, p'\} \in E$ is drawn as a polygonal chain of
	axis-aligned line segments (\emph{orthogonal polyline})
	that connects the drawings of~$p$ and~$p'$, and
	\item the total number of layers, the width of the drawing, the
	lengths of the edges, and the number of bends are
	kept reasonably small.
\end{itemize}

We have chosen this problem definition to be both simple and
extendable to more complex settings by using the described elements as building blocks.
For instance, if there are multiple edges per port,
then in a preprocessing we can assign each edge its own port and keep them together using a port group.
In a post-processing, we draw just one of these ports and we re-draw the ends of the edges incident to the other ports of this group.
Or if there are bundles of edges (e.g.\ a cable with twisted wires),
we can keep their ports together by introducing port groups.

Note that our problem definition generalizes the \textsc{Layered Graph
	Drawing} problem that is formalized and solved heuristically by the
Sugiyama framework~\cite{Sugiyama1981}.  Several subtasks of the
framework correspond to NP-hard optimization problems such as
\textsc{One-Sided Crossing Minimization}~\cite{ew-dg2l-TCS94}.  Hence,
we have to make do with a heuristic for our problem, too.  This
heuristic is coming up next.

\section{Algorithm}
\label{sec:algorithm}

\begin{figure}
	\begin{minipage}[b]{.36\textwidth}
		\begin{subfigure}[t]{\textwidth}
			\centering
			\includegraphics[]{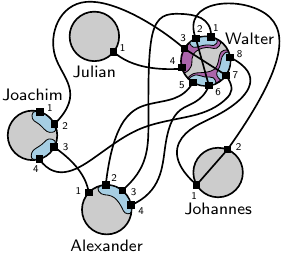}
			\caption{The input graph with five vertices.}
			\label{fig:full-example-step0}
		\end{subfigure}
		
		\bigskip
		
		\begin{subfigure}[t]{\textwidth}
			\centering
			\includegraphics[]{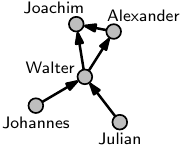}
			\caption{Phase 1: Orienting undirected edges using
				a force-directed graph drawing algorithm.}
			\label{fig:full-example-step1}
		\end{subfigure}
	\end{minipage}
	\hfill
	\begin{subfigure}[t]{.32\textwidth}
		\centering\includegraphics[scale=.75, trim=0 15 0 15, clip]{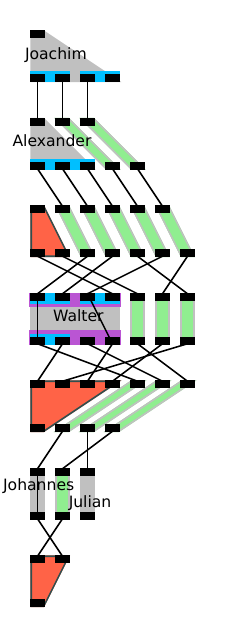}
		\caption{Phases 2 and 3: Assigning vertices to layers,
                  orienting ports and inserting dummy vertices.}
		\label{fig:full-example-step2}
	\end{subfigure}
	\hfill
	\begin{subfigure}[t]{.27\textwidth}
		\centering
		\includegraphics[scale=.75, trim=0 15 0 15, clip]{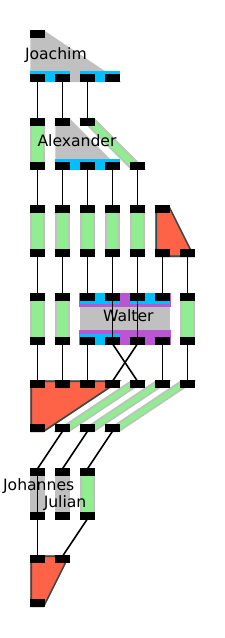}
		\caption{Phase 4: Reducing crossings by swapping vertices and ports.}
		\label{fig:full-example-step3}
	\end{subfigure}
	
	\bigskip
	
	\begin{subfigure}[t]{.22\textwidth}
		\centering
		\includegraphics[scale=.75, trim=10 15 10 15, clip]{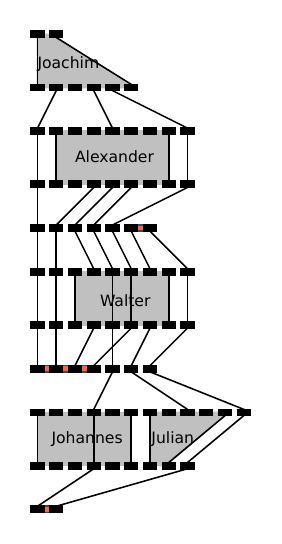}
		\caption{Phase 5.1: Transforming the drawing to a pure port structure.}
		\label{fig:full-example-step4}
	\end{subfigure}
	\hfill
	\begin{subfigure}[t]{.31\textwidth}
		\centering
		\includegraphics[scale=.75, trim=10 15 10 15, clip]{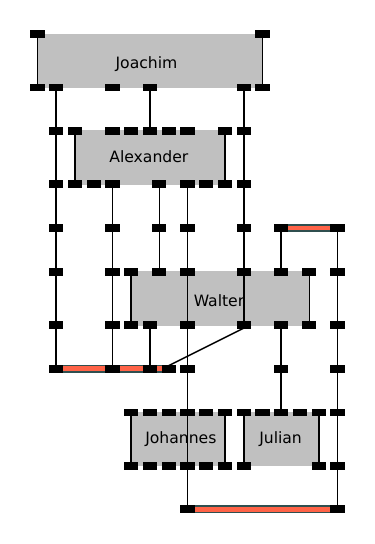}
		\caption{Phase 5.2: Determining vertex coordinates
			by aligning adjacent ports vertically.}
		\label{fig:full-example-step5}
	\end{subfigure}
	\hfill
	\begin{subfigure}[t]{.38\textwidth}
		\centering
		\includegraphics[scale=.75]{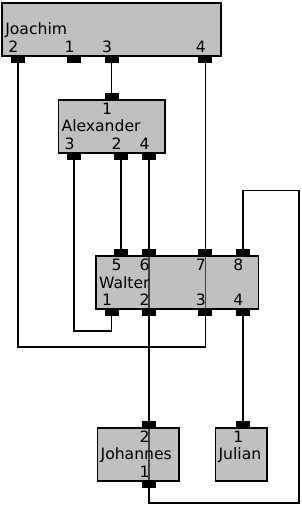}
		\caption{Phase 6: Constructing the drawing and routing the edges
			orthogonally. This is our final drawing.}
		\label{fig:full-example-step6}
	\end{subfigure}
	
	\caption{A full example that outlines how our algorithm works.
		Port groups are depicted in light blue and violet.
		(Vertex \textsf{Walter} has nested port groups.)
		Port pairings are indicated by straight-line segments
		inside vertices.
		Dummy vertices for long edges %spanning multiple layers
		are green; dummy vertices for turning edge directions
		on intermediate layers are red with a dark frame.}
	\label{fig:full-example}
\end{figure}

We assume that we are given a graph as described in \Cref{sec:preliminaries}.
(Otherwise we can preprocess accordingly.)
% the given data into the designated format.
Similarly to the algorithm of Sugiyama et al.~\cite{Sugiyama1981}, our
algorithm proceeds in the following phases, which we treat in the next subsections.
For a small but complete example, see \Cref{fig:full-example}.
\begin{description}[leftmargin=0pt] %, itemsep=-.15em]
	\item[Phase 1:] \emph{Orienting undirected edges.}
		 We orient the undirected edges by
		drawing the underlying simple graph with a force-directed 
		graph drawing algorithm and then direct all edges upwards.
		Alternatively, we may orient the edges by a breadth-first search
		in order of discovery.
		\hfill (\Cref{sub:orienting})
	\item[Phase 2:] \emph{Assigning vertices to layers.}
		\hfill (\Cref{sub:assigning})
	\item[Phase 3:] \emph{Orienting ports and inserting dummy vertices.}
		We try to place a port such that it is on the upper side of its vertex
		if its incident edge goes upwards and is on the lower side otherwise.
		However, due to port groups, port pairings and input constraints,
		a port may end up on the ``wrong'' side of its vertex. In this case,
		we subdivide the incident edge by a dummy vertex on a neighboring
		intermediate layer to turn the edge direction.
		\hfill (\Cref{sub:dummy-vertices})
	\item[Phase 4:] \emph{Reducing crossings by swapping vertices and ports.}
		We employ the classic barycenter heuristic by
		Sugiyama et al.~\cite{Sugiyama1981} on a port-wide level
		to reduce the number of edge crossing.
		\hfill (\Cref{sub:crossings})
	\item[Phase 5:] \emph{Determining vertex coordinates.}
		We transform our vertices to ports and apply
		the algorithm by Brandes and K\"opf~\cite{Brandes2001,Brandes2020}
		purly on the resulting port structure.
		\hfill (\Cref{sub:coordinates})
	\item[Phase 6:] \emph{Constructing the drawing.}  We resolve
		dummy ports and dummy vertices, and we route the edges
                orthogonally. \hfill (\Cref{sub:edge-routing})
\end{description}

\subsection{Orienting Undirected Edges}
\label{sub:orienting}
Classical algorithms for layered graph drawing expect as input a
directed acyclic graph, whose vertices are placed onto layers such
that all edges point downwards.
For directed cyclic graphs, some edges may be reversed or removed to make the graph acyclic.
In our case of undirected graphs, we suggest the following procedures
to orient the undirected edges, making the graph simultaneously
directed and acyclic.
(Hence, we don't need the cycle elimination phase of the Sugiyama framework.)
We ignore the ports in this step.
\begin{description}
	\item \bfs: We execute a breadth-first search from a random start
	vertex.
	Edges are oriented from vertices discovered earlier to vertices discovered later.
	\item \fd: We run a force-directed graph drawing algorithm.
	In the resulting drawing, edges are oriented upwards.
	% (or to the right if an edge ends up being perfectly horizontal).
	\item \rand: We place the vertices randomly into the drawing area,
	uniformly distributed.  In the resulting drawing, we orient the
	edges as in \fd.
\end{description}

The runtime of this phase is dominated by %the runtime of
the force-directed algorithm.
One might consider executing the force-directed algorithm more than once,
say $k$ times, with different random start positions and
then to use the drawing admitting the fewest crossings.
This is less time consuming than re-iterating the whole algorithm.
Note, however, that it is not clear whether a drawing with
fewer crossings is a much better starting point for the rest of the algorithm
and justifies the longer running time when choosing $k > 1$.
This question may be investigated in new experiments~--
we have always set $k = 1$.

In our experiments, we used a classical spring embedder~\cite{fr-gdfdp-SPE91}
with the speed-up technique as described by Lipp et al.~\cite{Lipp2016}.
The resulting runtime is in \mbox{$O(k \cdot I \cdot |V| \log |V|)$}, where $I$ is
the number of iterations per execution of the force-directed algorithm.

\subsection{Assigning Vertices to Layers}
\label{sub:assigning}

In this step we seek for an assignment of vertices to layers,
such that all directed edges point upwards.
We use a network simplex
algorithm as described by Gansner et al.~\cite{Gansner1993}.
The algorithm is optimal in the sense that the sum of layers the edges span is minimized.
With respect to the runtime of their algorithm, the authors state:
``Although its time complexity has not been proven polynomial, in
practice it takes few iterations and runs quickly.''
% s. auch wikipedia:
% https://en.wikipedia.org/wiki/Network\_simplex\_algorithm}

\subsection{Orienting Ports and Inserting Dummy Vertices}
\label{sub:dummy-vertices}

\begin{figure}[t]
	\hfill
	\begin{subfigure}[t]{.47\textwidth}
		\centering
		\includegraphics[page=2]{inserting-dummy-vertices}
		\caption{We insert an extra layer~$L_{2.5}$ to host a dummy
			vertex (solid red) as turning point.
			% for the two rightmost edges.
			All edges traversing a layer are subdivided by dummy
			vertices (hatched green).}
		\label{fig:inserting-dummy-vertices-after}
	\end{subfigure}
	\hfill
	\begin{subfigure}[t]{.47\textwidth}
		\centering
		\includegraphics[page=1]{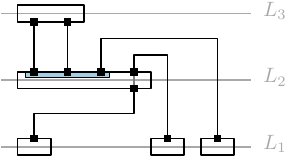}
		\caption{% Three vertices are assigned to different layers.
			Each port of the vertex on $L_2$ is in a port group or in a port
			pairing.  Thus, the two rightmost ports % of~$v_2$
			are placed on the top side, although they have
			incoming edges from below.}
		\label{fig:inserting-dummy-vertices-before}
	\end{subfigure}
	
	\caption{Example for the insertion of dummy vertices.}
	\label{fig:inserting-dummy-vertices}
\end{figure}

%An example for this step is depicted in Fig.~\ref{fig:inserting-dummy-vertices}.
Consider the ports of a vertex.
% If there is a port group with ports of type \ttop and \bbottom
% or a port pairing with twice \ttop or twice \bbottom,
% we report that there is no solution.
If a port group is of a type different than \ffree,
we assign all ports of this port group or a port group containing this port group
to the specified vertex side, e.g., the bottom side.
(Ignore for the moment the port groups of type \lleft and \rright.
Below, we describe how to handle them.)
If this leads to contradicting assignments of the same port,
then the input is inconsistent in assigning vertex sides to ports.
We arbitrarily change vertex sides of affected port groups to obtain consistency.
(Alternatively, one could reject such an instance.)
We treat port pairings analogously.
We assign ports that are in no port group to the top or the bottom side
depending on whether they have an outgoing or incoming edge.
If ports of a port group of type \ffree remain unassigned,
we make a majority decision for the top-level port group---if there are more outgoing than incoming
edges, we set its ports to the top side; otherwise to the bottom side.

In any case, we may end up with ports being on the ``wrong'' side
in terms of incident edges,
e.g., a port on the top side has an incoming edge.
To make such edges reach their other endpoints without running through
the vertex rectangle,
we introduce an extra layer directly above the layer at hand.
On the extra layer, we then place a dummy vertex that will serve as a
``turning point'' for these edges;
see Fig.~\ref{fig:inserting-dummy-vertices}.
We will refer to them as \emph{turning dummy vertices}.

%
% Instead of using dummy vertices on an extra layer, 
In contrast, \kieler~\cite{ssh-dlgpc-JVLC14} appends effectively, for each port
that lies on the ``wrong'' side, a dummy port on the opposite side of
the vertex rectangle, to the very right or left of the ports there.
The edges will later be routed around the vertex to this dummy port.
Our new approach therefore provides a somewhat greater flexibility in
routing edges around vertices.

\begin{figure}[t]
	\begin{subfigure}[t]{.47\textwidth}
		\centering
		\includegraphics[page=1]{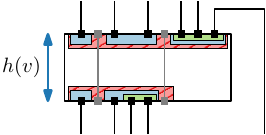}
		\caption{Instead of ports on the left and the right side,
			we subdivide the top and bottom side into three port groups (solid blue) using
			a port group with fixed order (hatched red) and two port pairings.}
		\label{fig:left-right-side-ports-before}
	\end{subfigure}
	\hfill
	\begin{subfigure}[t]{.47\textwidth}
		\includegraphics[page=2]{left-right-side-ports}
		\caption{In a post-processing, we shrink a vertex to its middle part and re-route the edges entering a port on the left or right side of the vertex.
			The considered vertex has two port groups (solid green).}
		\label{fig:left-right-side-ports-after}
	\end{subfigure}
	
	\caption{Construction to model ports on the left and the right side of a vertex.}
	\label{fig:left-right-side-ports}
\end{figure}

It remains to describe how to handle port groups of type \lleft and \rright.
Note that our algorithm never assigns ports of a port group of type \ffree
to \lleft or \rright.
However, the input data may contain port groups of these types\footnote{
In our experiments, we do not have port groups of type \lleft or \rright.
So here we suggest a general approach how to handle this case,
which we did not implement or test.}.
Consider the port groups of type \lleft and \rright;
see Fig.~\ref{fig:left-right-side-ports} for this step.
We assign their ports during the execution of the algorithm
to the bottom or the top side of their
vertices---again by a majority decision on their top-level port group.
On the top and the bottom side, we introduce new
top-level port groups with fixed order (hatched red in Fig.~\ref{fig:left-right-side-ports-before}).
They contain three port groups of free order (solid blue in Fig.~\ref{fig:left-right-side-ports-before}) that contain
everything on the left side, top/bottom side, and right side
(in this order and each separated by two ports with a port pairing; gray in Fig.~\ref{fig:left-right-side-ports-before}).
Later, we will shrink each vertex~$v$ to its inner part and
re-route the ends of the edges incident to ports in port groups of type~\lleft and~\rright
as L-shapes in the released area (interior of the dashed box in Fig.~\ref{fig:left-right-side-ports-after}).
Hence, we adjust $w(v)$ and $h(v)$ in the forehand accordingly.

After this step for handling port groups of type \lleft and \rright,
every port is assigned either to the top or the bottom side of its vertex.

As in the classical algorithms for layered graph drawing, we subdivide
edges traversing a layer (which may also be an extra layer) by a new
dummy vertex on each such layer.  Hence, we have only edges connecting
neighboring layers.
As for all algorithms that rely on decomposing the edges, this phase
runs in time $O(\lambda \cdot |E| + |P|)$, where $\lambda$ is the number of
layers.  Note that $\lambda \in O(|V|)$.

\subsection{Reducing Crossings by Swapping Vertices and Ports}
\label{sub:crossings}

We employ the layer sweep algorithm using the well-known barycenter
heuristic proposed by Sugiyama et al.~\cite{Sugiyama1981}.
However, we also have to take the ports and the port constraints into
account.
We suggest three ways to incorporate them.

\begin{description}
	\item \vertices: We first ignore ports.  We arrange the vertices as follows.
	Since there may be many edges between the same pair of vertices,
	we compute the vertex barycenters weighted by edge multiplicities.
	After having arranged all vertices, we arrange the ports at each vertex to minimize edge crossings.
	Finally, we rearrange the ports according to port pairings and port groups by computing barycenters of the ports of each port group.
	\item \ports: We use indices for the ports instead of the vertices
	and apply the barycenter heuristic to the ports.
	This may yield an invalid ordering with respect to port groups and vertices.
	Hence, we sort the vertices by the arithmetic mean of the port
	indices computed before.
	Within a vertex, we sort the port groups by the arithmetic mean of the indices of their ports.
	We recursively proceed in this way for port groups contained in port groups and finally for the ports.
	\item 	\mixed: Vertices that do not have port pairings are kept as a whole,
	vertices with port pairings are decomposed into their ports.
	The idea is that, when sweeping up or down, the ports do not
	influence the ordering on the other side
	and can be handled in the end---unless they are paired.
	After each iteration, we force the ports from decomposed
	vertices to be neighbors by computing their barycenters,
	and we arrange the paired ports above each other.
	Finally, we arrange all ports that are not included in the ordering as in \vertices.
\end{description}
In all cases, if a port group has fixed order, we cannot re-permute
its elements, but we take the order as described from left to right.
We use random start permutations for vertices and ports.
We execute this step $r$ times for some constant~$r$ (in our experiments $r=1$) and take the solution that causes
the fewest crossings.

\kieler~\cite{ssh-dlgpc-JVLC14} also computes barycenters depending on the order of ports of the previous layer.
Similar to \ports they describe a \emph{layer-total} approach and
similar to \mixed they describe a \emph{node-relative} approach.
However, they compute barycenters only for vertices as a whole.
We use barycenters of ports to recursively determine also an ordering of port groups.

It remains to describe how to handle a vertex $v$ on a layer $L_i$
that has edges in only one direction, say to the layer~$L_{i-1}$ below.
In particular, this concerns turning dummy vertices
of which we have many in our experiments.
If we sweep upwards, we use $v$'s neighbors on $L_{i-1}$
to determine $v$'s barycenter $b_{v^-}$
in the usual way,
which is $$b_{v^-} = \frac{\sum_{u \in N(v) \cap L_{i-1}} \textrm{pos}_{L_{i-1}}(u)}{|N(v) \cap L_{i-1}|} \, ,$$
where $N(v)$ are $v$'s neighbors and $\textrm{pos}_{L_{i-1}}(u)$ is the
position of vertex $u$ on layer $L_{i-1}$.
However, if we sweep downwards, it is not clear how to
arrange $v$ relative to the other vertices on $L_{i}$
since we cannot compute a barycenter using neighboring
vertices on $L_{i+1}$.

For these local sources and sinks, we investigate the following strategies\footnote{%
In the conference version~\cite{Walter2020a} of this article, we only used \pseudobc.}.
\begin{description}
	\item \pseudobc: We compute and use a pseudo
	barycenter~$b_{v^+}^{\textrm{pseudo}}$ being the current position of $v$
	on its layer $L_i$ normalized by the number of vertices on $L_{i + 1}$.
	More precisely, $b_{v^+}^{\textrm{pseudo}} = \textrm{pos}_{L_{i}}(v) \cdot \frac{|L_{i + 1}|}{|L_{i}|}$.
	\item \oppositebc: We compute and use a barycenter~$b_{v^+}^{\textrm{opposite}}$
	being the barycenter of $v$ with respect to the opposite layer of $L_i$
	normalized by the number of vertices on $L_{i + 1}$.
	More precisely, $b_{v^+}^{\textrm{opposite}} = b_{v^-} \cdot \frac{|L_{i + 1}|}{|L_{i-1}|}$.
	\item \relpos: We do not compute any barycenter of $v$,
	but keep $v$ at its current position within $L_i$.
	In other words, we remove $v$ and all vertices without edges to $L_{i+1}$ 
	from $L_i$ before computing the barycenters.
	Then, we sort the remaining vertices in the usual way according to their
	barycenters with respect to $L_{i+1}$.
	Finally, we re-insert $v$ and all vertices without edges to $L_{i+1}$
	into the same positions they previously had on $L_i$.
\end{description}

This phase runs in time $O(r \cdot J \cdot \lambda \cdot |E|)$, 
where $J$ is the number of % top-down 
(top-down or bottom-up) sweeps within one execution of the layer sweep algorithm.

\subsection{Determining Vertex Coordinates}
\label{sub:coordinates}

\begin{figure}[t]
	\centering
	\begin{subfigure}[t]{.46\textwidth}
		\centering
		\includegraphics[page=1]{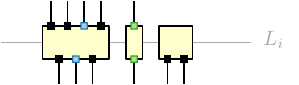}
		\caption{three vertices with two port pairings on one layer before transforming them to ports only}
		\label{fig:transform-vertices-to-ports-before}
	\end{subfigure}
	\hfill
	\begin{subfigure}[t]{.46\textwidth}
		\centering
		\includegraphics[page=3]{transform-vertices-to-ports}
		\caption{only ports on two layers; port pairings are
		connected by a dummy edge, the rightmost vertex is
		``padded'' to be wider using dummy ports
		}
		\label{fig:transform-vertices-to-ports-after}
	\end{subfigure}
	\caption{Example of the transformation of vertices with ports on one layer to ports and edges on two layers;
		port pairings are indicated by color.}
	\label{fig:transform-vertices-to-ports}
\end{figure}

To position both vertices and ports, we decompose the vertices into ports and edges.
An example is given in Fig.~\ref{fig:transform-vertices-to-ports}.
We duplicate each layer~$L_i$ (except for the extra layers introduced
in Section~\ref{sub:dummy-vertices}) to an
upper layer~$L_{i^+}$ and a lower layer~$L_{i^-}$.
For a vertex on layer~$L_i$, we place all ports of the \ttop side in the previously computed order onto~$L_{i^+}$ and all ports of the \bbottom side in the previously computed order onto~$L_{i^-}$.
To separate the vertices from each other and to assign them a
rectangular drawing area, we insert a path of length one with the one
port on~$L_{i^-}$ and the other port on~$L_{i^+}$ at the beginning and
the end of each layer and between every two consecutive vertices
(gray with ports drawn as disks in Fig.~\ref{fig:transform-vertices-to-ports}(b)).
Moreover, we may insert dummy ports without edges within the
designated area of a vertex, to increase the width of a vertex.
This can be seen as ``padding'' the width of a vertex~$v$ via ports to obtain the desired minimum width~$w(v)$.
For each port pairing $\{p, p'\}$, where $p$ is on~$L_{i^-}$ and $p'$ is on~$L_{i^+}$,
we insert a dummy edge connecting $p$ and~$p'$.
Similarly for each dummy vertex subdividing a long edge,
we add a path of length~1 between~$L_{i^-}$ and~$L_{i^+}$.
Observe that we do not have edge crossings between~$L_{i^-}$ and~$L_{i^+}$.
Therefore, using the algorithm of Brandes and K{\"o}pf~\cite{Brandes2001}
(see below), these edges will end up as vertical line segments.
This fulfills our requirement for vertices being rectangular and for
ports of port pairings being vertically aligned.

Now we have a new graph~$G'$ with ports being assigned to layers, but without vertices and without port constraints.
So, in the following we consider the ports as vertices.
This is precisely the situation as in the classical algorithms
for layered graph drawing when determining coordinates of vertices.
After the current coordinate assignment step, we will re-transform the
drawing into our setting with vertices, ports, and edges.

The y-coordinate of a port is given by its layer.
For assigning x-coordinates, we use the well-established linear-time algorithm of Brandes and K{\"o}pf~\cite{Brandes2001}.
It heuristically tries to straighten long edges vertically and balancing the position of
a port with respect to its upper and lower neighbors.
It guarantees to preserve the given port order on each layer and a
minimum distance~$\delta$ between consecutive ports.
Moreover, it guarantees that uncrossed edges are drawn as vertical
line segments, which is crucial for our application.
Such a sequence of vertically stacked ports is called a \emph{block}.
Roughly speaking, the blocks are placed horizontally next to each other
such that no two blocks overlap and the slack between the blocks is minimized.

We note that the original algorithm of Brandes and
K{\"o}pf~\cite{Brandes2001} contained two flaws that came up in our
experiments.  Subsequently, they were fixed~\cite{Brandes2020}.

\begin{figure}[t]
	\centering
	\begin{subfigure}[t]{.46\textwidth}
		\centering
		\includegraphics[page=1]{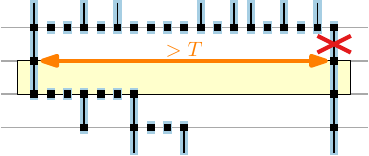}
		\caption{finding a large gap within a vertex}
		\label{fig:tighten-nodes-before}
	\end{subfigure}
	\hfill
	\begin{subfigure}[t]{.50\textwidth}
		\includegraphics[page=2]{tighten-nodes}
		\caption{breaking a block into two parts to narrow the gap}
		\label{fig:tighten-nodes-after}
	\end{subfigure}
	\caption{Example of a wide vertex (yellow background color) arising during the execution
		of the algorithm of Brandes and K{\"o}pf~\cite{Brandes2001}.
		With an additional check, we detect large gaps between neighboring
		ports within a vertex and ``break'' the involved blocks.
		Here, blocks are highlighted by blue background color.}
	\label{fig:tighten-nodes}
\end{figure}

Using the algorithm of Brandes and K{\"o}pf
for ports instead of vertices has the drawback that
vertices are drawn as relatively wide rectangles.
This is because ports of the same vertex may be placed vertically
above distant ports of the previous layer.
To avoid these large gaps between ports of the same vertex,
we extend the algorithm of Brandes and K{\"o}pf by the following
check when placing the blocks.
If two ports of two neighboring blocks are part of the same vertex 
and if the distance between these two ports is greater than a given threshold~$T$
(in our case 16 times the given minimum port distance),
then we ``break'' one of the involved blocks into two blocks;
see \Cref{fig:tighten-nodes}.
This means that one of the edges that has been a vertical edge within the
block is not drawn as a vertical line segment.
However, now the blocks are placed closer to each other
effecting a smaller total width of the vertex.

It may happen that a large gap cannot be closed this way
because we are not allowed to break port pairing edges.
Therefore, we additionally do a post processing, 
where we forget about all blocks and structures within
the algorithm of Brandes and K{\"o}pf and just consider
each vertex individually. If large gaps remain,
we push ports closer to each other where possible
without breaking internal port pairings.
Note that by avoiding wide vertices with both of these operation,
we increase the number of bends in the resulting drawing
since we lose vertical straight-line segments.

The algorithm of Brandes and K{\"o}pf runs in time linear in the number of ports and edges.
Our modification breaks each block at most $\lambda$ times,
where $\lambda$ is the number of layers.
Hence, this phase runs in time $O(\lambda (|E| + |P|))$. \todo{JZ: I think this estimation is not tight, but at least it should be correct. I think it is not worth the effort trying to tighten it (this might require a much more detailed description) unless there is a simple argument.}

\subsection{Constructing the Drawing and Routing the Edges Orthogonally}
\label{sub:edge-routing}

First, we obtain vertices drawn as rectangles from (dummy) ports and
edges by reversing the transformation described in
\Cref{sub:coordinates}.

Then, we  obtain edges drawn as polylines by
transforming the dummy vertices inserted in \Cref{sub:dummy-vertices}
into bend points of their edges.
We re-draw vertices with ports on the left or right side
by shrinking the width of the vertex and extending the incident
edges within the released area.
For horizontal port pairings, we increase the height of a vertex
and re-sort the ports on the left and the right side.

Finally, we draw the edges orthogonally.
We describe this in more detail in the remainder of this section.

\begin{figure}[t]
	\hfill
	\begin{subfigure}[t]{.47\textwidth}
		\centering
		\includegraphics[page=1]{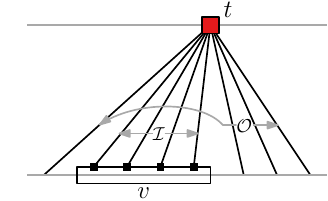}
		\caption{initial situation}
		\label{fig:resolve-turning-dummies-before}
	\end{subfigure}
	\hfill
	\begin{subfigure}[t]{.47\textwidth}
		\centering
		\includegraphics[page=2]{resolve-turning-dummies}
		\caption{drawing stacked ``U''s}
		\label{fig:resolve-turning-dummies-after}
	\end{subfigure}
	
	\caption{Drawing edges going through turning dummy vertices orthogonally.}
	\label{fig:resolve-turning-dummies}
\end{figure}

Here, let us first describe how to draw the edges going through a turning dummy vertex~$t$
(red in Fig.~\ref{fig:inserting-dummy-vertices}).
This step is depicted in Fig.~\ref{fig:resolve-turning-dummies}.
Recall that for each vertex~$v$, we have up to one turning dummy vertex
on the next layer above (for edges going downwards)
and up to one in the next layer below (for edges going upwards).
Without loss of generality, let~$t$ be on the next layer above~$v$.
Observe that we have an even number of edge pieces being
adjacent to~$t$ as they correspond to edges entering
and leaving~$t$.
Let~$\mathcal{I}$ be the set of edge pieces entering~$t$, and
let~$\mathcal{O}$ be the set of those leaving~$t$.
Those in~$\mathcal{I}$ are incident to ports~$P_\mathcal{I}$ of~$v$.
Where possible, we sort the ports of $P_\mathcal{I}$ at~$v$ such that
the order of $\mathcal{I}$ is, for both the edges passing~$v$ on the left and
on the right, inverse to their corresponding edge pieces in~$\mathcal{O}$.
This can be done in time~$O(\lambda |E|)$ in total using Bucketsort.
The resulting order allows us to draw the edges as two stacks of
(upside-down) ``U''s as in Fig.~\ref{fig:resolve-turning-dummies-after}.
We greedily use intermediate lines $\ell_1, \ell_2, \dots$ to place the
horizontal pieces. % OR: on which we place the horizontal pieces.
Since we need at most $O(|E|)$ lines between any two layers
and have at most~$O(\lambda |E|)$ edge pieces, the runtime for this step is
$O(\lambda |E|^2)$ in the worst case.
The greedy procedure is optimal for an individual vertex, but may
produce avoidable crossings between different vertices depending on
the order in which we process the dummy turning vertices.

For all other edge pieces spanning a layer, it remains to draw them orthogonally.
We do not need to consider vertical segments since they are already
drawn in the orthogonal style.
% Moreover, we ignore the edge pieces we have just drawn as ``U''s for now.
% Recall that most of these ``edges'' are actually pieces of longer edges.
Consider the remaining (skewed) edge pieces.  Since they are
directed upwards, we will refer to them as \emph{arcs} (with arc set
$A$).  Their endpoints are ports of vertices and dummy vertices.
Let $P$ be the set of these ports.  (This ignores ports of degree~0).
% For simplicity,
We first assume that the x-coordinates of the ports on the
two layers are all different.  Below, we treat the general case.

%Note that
The graph $M=(P,A)$ is a perfect matching.  Each port $u \in
P$ has its x-coordinate $x(u)$.  For an arc $uv$, $\xspan(uv) = [\min
\{x(u),x(v)\}, \max \{x(u),x(v)\}]$ is its \emph{span}.  
We have two types of arcs; $uv$
is \emph{right-going} if $x(u)<x(v)$ and \emph{left-going} otherwise.
We want to draw each arc $uv$ as a sequence of three axis-aligned line
segments: vertical, horizontal, vertical; starting at $u$ and ending
at~$v$.  For the horizontal pieces we use horizontal lines.  Our task
is to assign the horizontal piece of each arc~$a$ to a line
$\xline(a)$ such that no two horizontal pieces intersect and such that
the number of lines is minimized.

Without further restrictions, this would correspond to partitioning
the set $\{ \xspan(a) \colon a \in A \}$ into as few independent sets
as possible.  We require, however, that every pair of arcs intersects
at most once.  For two right-going arcs $uv$ and $u'v'$ with
$x(u)<x(u')<x(v)<x(v')$, this implies that $\xline(uv)>\xline(u'v')$.
Symmetrically, for two left-going arcs $uv$ and $u'v'$ with
$x(v)<x(v')<x(u)<x(u')$, this implies that $\xline(uv)<\xline(u'v')$.

\begin{figure}[t]
	\begin{minipage}[b]{.56\textwidth}
		\centering
		\includegraphics{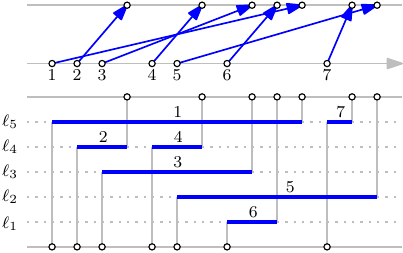}
		\caption{Drawing right-going arcs.}
		\label{fig:edge-routing}
	\end{minipage}
	\hfill
	\begin{minipage}[b]{.40\textwidth}
		\centering
		\includegraphics{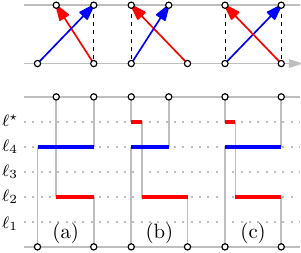}
		\caption{Equal x-coordinates.}
		\label{fig:special-routing}
	\end{minipage}
\end{figure}

We solve this combinatorial optimization problem as
follows.  We first go through the left-going arcs in the left-to-right
order of their upper endpoints.  We place each arc greedily
on the lowest available line.  Then we solve the problem for the
right-going arcs symmetrically, in the left-to-right order of their
lower endpoints, placing them on the highest available line; see
Fig.~\ref{fig:edge-routing}.
Again, this can be accomplished in time~$O(\lambda |E|^2)$ in the worst case.

If, for the left- and right-going arcs, there are
ports with equal x-coordinates (connected by black dashed
lines in Fig.~\ref{fig:special-routing}, top row), we must additionally make
sure that their vertical segments don't intersect.  To this end, we
introduce an additional line~$\ell^\star$ at the top to place an extra
horizontal segment for all ``problematic'' cases, investing two
additional bends; see Fig.~\ref{fig:special-routing}(b) and~(c).  In
Fig.~\ref{fig:special-routing}(a) (where the right endpoints have the
same x-coordinate) no extra bends are needed because we place the
left-going arcs below the right-going arcs.

\begin{figure}[t]
	\centering
	\includegraphics{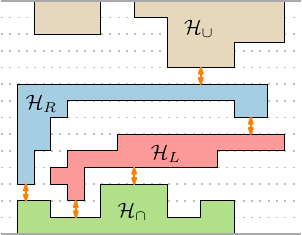}	
	\caption{Moving the horizontal pieces of the ``U''s, the right-going arcs,
		the left-going arcs, and the upside-down ``U''s towards each other.}
	\label{fig:moving-contour-lines}
\end{figure}

Finally, we move the horizontal pieces~$\mathcal{H}_R$ of right-going arcs simultaneously
down until at least one of these pieces, say $a$, is only one line
above a horizontal piece, say~$b$, (which is in the group of
horizontal pieces~$\mathcal{H}_L$ of left-going arcs) with
$\xspan(a)\cap\xspan(b)\ne\emptyset$.
We do the same for the ``U''s ($\mathcal{H}_\cup$) on the top and the
upside-down ``U''s ($\mathcal{H}_\cap$) on the bottom.
In other words, we move the blocks of horizontal edge pieces
towards each other until their contour lines would overlap
if we would move by another line; see Fig.~\ref{fig:moving-contour-lines}.

The greedy approach for placing only left-going (right-going) arcs
is optimal in terms of the number of used lines~\cite{Mittelstaedt2022},
which makes our merging of left-going and right-going
arcs a 2-approximation.

It remains to analyze the running time of this step.
Between each two layers, we can merge all contour points into a list
in time~$O(|E|)$ and then use a sweep-line approach to determine the
distances between the contour lines between each to points of the list---again in time~$O(|E|)$.
So over all layers, this step can be performed in time~$O(\lambda |E|)$.

The total runtime of this phase is $O(\lambda \cdot |E|^2)$ in the worst case
(while in practice we would rather expect a linear %than a quadratic
runtime behavior).

\section{Experimental Evaluation}
\label{sec:evaluation}

For our experiments we got access to 380 real cable plans of a large
German machine manufacturing company (and another smaller data set;
see \Cref{sub:experiments}). To obfuscate these plans and to
have more data for our experiments, we generated 1140 pseudo cable
plans from the real cable plans---three from each real cable plan.
For replicability, we have made all of our algorithms,
data structures, and data described here publicly available on
github~\cite{github-praline,github-pseudo-plans}---except for the
original (company-owned)~plans.

\subsection{Graphs Used in the Experiments}
First, we discuss the structure of these cable plans and how we
transformed them to the format that is expected by our algorithm.
% This can be seen as an example of how to transform real-world graphs
% with specific port constraints to our format.
A cable plan has vertices with ports and vertex groups that comprise multiple vertices.
Moreover, there can be edges connecting two or more ports (that is,
hyperedges) and a port can be incident to an arbitrary number of
edges.
In a vertex group, there are port pairings between two vertices and these vertices should be drawn as touching rectangles.
In our model, we do not have vertex groups and port pairings between different vertices.
Instead, we model a vertex group as a single vertex with (internal)
port pairings and a port group for the ports of each vertex.
Moreover, we split ports of degree~$d$ into $d$ separate ports and
enforce that they are drawn next to each other and on the same side of
the vertex by an (unordered) port group.
We replace hyperedges by a dummy vertex having an edge to each of the ports of the hyperedge.
We don't have ports on the left or the right side of a vertex.

\subsection{Generating a Large Pseudo Data Set from Original Data}
\label{sub:generating}

\begin{figure}[t]
	\begin{subfigure}[t]{.42\textwidth}
		\centering
		\includegraphics[scale=.6]{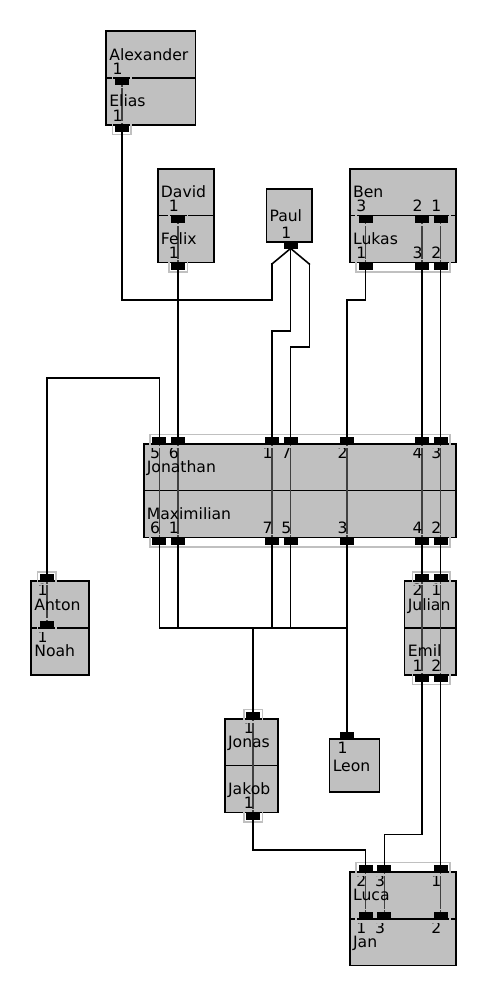}
		\caption{anonymized original cable plan}
		\label{fig:original-plan}
	\end{subfigure}
	\hfill
	\begin{subfigure}[t]{.54\textwidth}
		\centering
		\includegraphics[scale=.6]{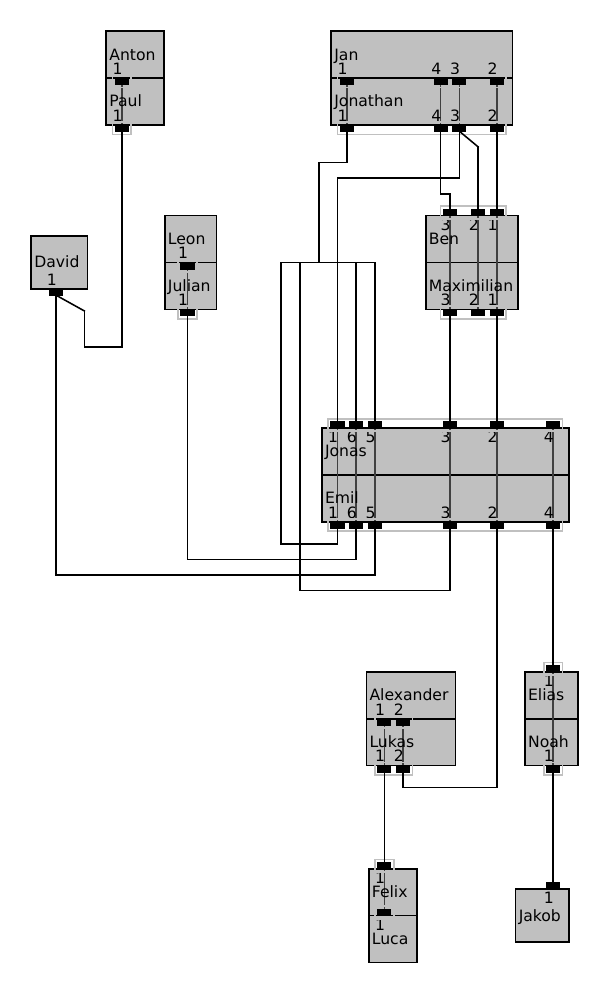}
		\caption{artificial cable plan generated from the plan in~(\subref{fig:original-plan})}
		\label{fig:pseudo-plan}
	\end{subfigure}
	
	\caption{Example of an artificial cable plan generated from an original cable plan.
		Port groups are indicated by gray boxes and port pairings by line segments inside a vertex.}
	\label{fig:original-pseudo-plan}
\end{figure}

Now, we describe briefly how we generated the pseudo cable plans.
This can be seen as a method to extend and disguise a set of real-world graphs.
A drawing of an original cable plan and a derived pseudo cable plan is depicted
in Fig.~\ref{fig:original-pseudo-plan}.
In \ref{app:drawings},\todo{AW: The gap between ``Appendix'' and ``A''
  is too large, but never mind for now.} we
show larger examples of drawings of original cable plans and pseudo cable plans.
We generate a pseudo plan by removing and inserting elements from/to an original plan.
Elements of the plans are the vertex groups, vertices, ports, port pairings, and edges.
As a requirement we had to replace or remove at least a $q$-fraction
of the original elements (in our case $q = .05$).
We proceed in three phases.
\begin{enumerate}
	\item We determine target values for most elements of the graph (number of vertex groups, vertices, ports, port pairings)
	and more specific parameters (distribution of edge--port incidences,
	arithmetic mean of parallel edges per edge, number of self loops,
	distribution of ports per edge, distribution of edges per port).
	We pick each target value randomly using a normal distribution,
	where the mean is this value in the original plan
	and the standard deviation is the standard deviation of this value across all graphs of the original data set divided by the number of plans in the original data set times a constant.
	\item We remove a $q$-fraction of the original elements uniformly at random in the following order:
	vertex groups (incl.\ contained vertices and incident edges), vertices (incl.\ ports and incident edges), port pairings (incl.\ ports and incident edges), ports (incl.\ incident edges), and edges.
	\item In the same order, we add as many new elements as needed to reach the respective target values.
	For the insertion of edges we are a bit more careful.
	In case the graph became disconnected during the deletion phase,
	we first reconnect the graph by connecting different components.
	Then, we insert the remaining edges according to the distributions of
	edge--port incidences while trying to reduce the gaps between the target value
	and the current value for parallel edges per edge and for the number of self loops.
	Parallel edges have the same terminal vertices but not necessarily the same terminal ports.
	We mostly use ports that do not have edges (they are new or their edges were removed or they had no edges initially)
	and assign for each one the number of edges it should get in the end.
	This gives us a set of candidate ports.
	Next, we iteratively add a (hyper)edge~$e$ connecting $d$ ports.
	In each iteration,
	we pick $c$~sets of $d$~ports from our set of candidate ports uniformly at random---each set is a candidate for the end points of the new edge.
	We choose the set where we approach the aforementioned target values
	the best if we would add the corresponding edge to the current graph.
	We used $c = 1000$, which means we took one out of 1000 randomly generated edge candidates.
\end{enumerate}

Our generated pseudo cable plans are good if they are similar to and have similar characteristics as the original cable plans,
and if the corresponding original cable plans cannot easily be reconstructed from the pseudo cable plans.

For our purposes, we can compare the results of the experiments using the original data set and the generated data set or
we can compute explicit graph characterization parameters.
The numbers of vertices, ports, edges, etc.\ are similar by using the target values.
For example,
the arithmetic mean (median) of the number of vertices in the original data set is 106.21 (106),
while it is 106.15 (105.5) in the generated data set.
The arithmetic mean (median) across the arithmetic means of parallel edges per edge in the original data set is 1.590 (1.429),
while it is 1.491 (1.401) in the generated data set.
Some characteristic parameters where we did not have target values exhibit at least some similarities,
which indicates a similar structure of the graphs of both sets.
For example,
the arithmetic mean (median) of the diameters across the largest components of all graphs in the original data set is 9.508~(10),
while it is 8.731~(9) in the generated data set.

\subsection{Experiments}
\label{sub:experiments}
Our experiments were run in Java on an Intel Core~i7 notebook with 8 cores (used in parallel) and 24 GB RAM under Linux and took about 3~hours.

We note that we have another smaller data set of 192 real cable plans
where the vertex labels are common German male given names.
We call this data set \emph{readable data set}
and the previously described data set \emph{large data set}.
From the readable data set, we have generated pseudo cable plans as well.
As it turned out, the statistical results for both data sets
are very similar.  This supports the stability of our results.
Due to the similarity of the results, we decided to detail only
the results of the large data set in the description of our experiments.
However, we present drawings of both data sets in \ref{app:drawings},
and the generated pseudo plans of both data sets are available
in the git repository~\cite{github-pseudo-plans}.

\subsubsection{Orienting Undirected Edges}
For each graph %of the 1135 graphs
and each of the variants \fd, \bfs, \rand, we oriented the edges and
executed the algorithm ten times
using the variant \ports in the crossing reduction phase.
%with a single repetition. \todo{AW: Also 2x?
%JZ: no, just once. I've removed this to avoid confusions.}
For \fd, we used only one execution of the force-directed algorithm (so $k = 1$)
to make it better comparable to the other methods.  We recorded
\begin{itemize}
	\item the number~\ncr of crossings in the final drawing,
	\item the number~\nbp of bends created when executing the algorithm,
	\item the width, height, total area, and aspect ratio of the
          bounding box of the drawing, and
	\item the time to orient the edges and run the algorithm.
\end{itemize}

\begin{table}[tbh]
	\setlength{\tabcolsep}{4pt}
	\caption{Comparison of the methods for orienting the edges.
		The mean $\mu$ is relative to \rand (standard deviation in the
		range $[.1, .3]$); $\beta$ measures (in \%) how often
		a method provides the best result ($\sum \beta>100$ possible
		due to ties).}
	\label{table:results-orienting}
	
	\medskip
	
	\centering
	
	\begin{tabular}{@{}r@{~~}|@{~~}rr@{~~~~~}rr@{~~~~~}rr@{~~~}|@{~~~}rr@{~~~~~}rr@{~~~~~}rr@{}}
		\toprule
		\multicolumn{1}{c}{} & 
		\multicolumn{6}{c@{~~~~}}{original cable plans} & 
		\multicolumn{6}{c}{generated artificial cable plans}
		\\ \midrule
		\multicolumn{1}{c}{} &
		\multicolumn{2}{c@{~~~}}{\fd} &
		\multicolumn{2}{c@{~~~}}{\bfs} &
		\multicolumn{2}{c@{~~~}|@{~~~}}{\rand} &
		\multicolumn{2}{c@{~~~}}{\fd} &
		\multicolumn{2}{c@{~~~}}{\bfs} &
		\multicolumn{2}{c}{\rand} \\
		& $\mu$ & $\beta$ & $\mu$ & $\beta$ & $\mu$ & $\beta$
		& $\mu$ & $\beta$ & $\mu$ & $\beta$ & $\mu$ & $\beta$ \\
		\midrule
		\ncr & ~.57 & \textbf{78} & ~.64 & 31 & 1 & 11
		& ~.66 & \textbf{87} & ~.76 & 25 & 1 & 11 \\
		\nbp & ~.94 & \textbf{68} & ~.96 & 33 & 1 & 15
		& ~.96 & \textbf{75} & ~.99 & 22 & 1 & 17 \\
		width & ~.56 & \textbf{92} & ~.75 & 12 & 1 & 2
		& ~.64 & \textbf{93} & ~.80 & 9 & 1 & 2 \\
		height & 1.80 & 3 & 1.42 & 4 & 1 & \textbf{97}
		& 1.69 & 1 & 1.37 & 4 & 1 & \textbf{98} \\
		area & ~.98 & \textbf{54} & 1.04 & 18 & 1 & 33
		& 1.06 & 27 & 1.06 & 26 & 1 & \textbf{50} \\
		w:h & ~.49 & \textbf{85} & ~.65 & 17 & 1 & 3
		& ~.56 & \textbf{86} & ~.73 & 13 & 1 & 5 \\
		time & 1.10 & 3 & ~.81 & \textbf{97} & 1 & 11
		& 1.17 & 2 & ~.87 & \textbf{90} & 1 & 19 \\
		\bottomrule
	\end{tabular}
\end{table}
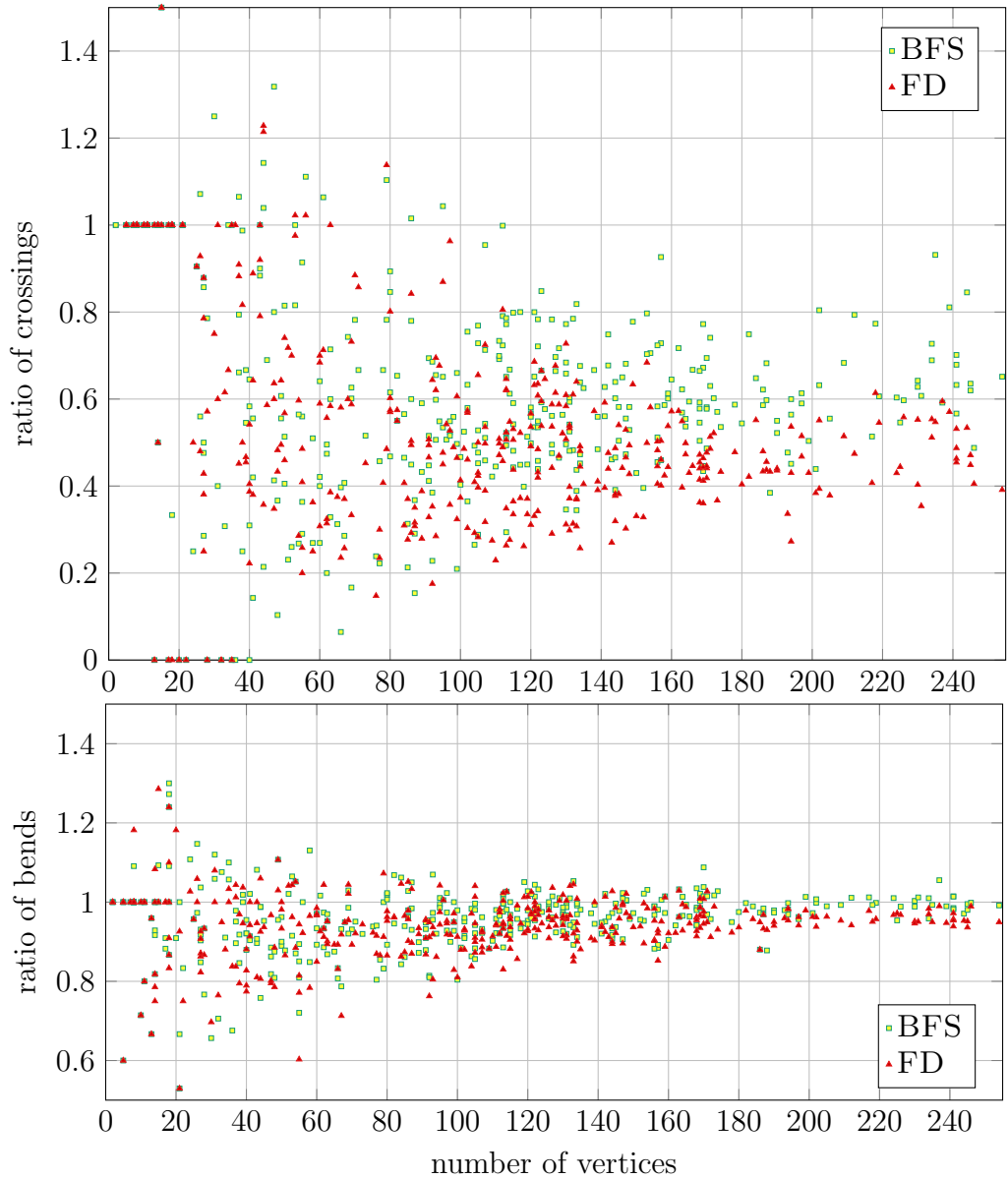
\begin{figure}[p]
	\centering
	\begin{tikzpicture}
	\begin{axis}[
	width=\textwidth,
	height=.54\textheight,
	legend pos=north east,
	legend cell align={left},
	xmin=0,
	xmax=255,
	ymin=0,
	ymax=1.5,
	grid=both,
	ylabel={ratio of crossings},]
	
	\addplot+[only marks, color=cb-Dark2-1,  every mark/.append style={solid, fill=cb-Set1-6, scale=.45}, mark=square*]
	table[x=vtcs,y=bfs/rand]
	{ncrDA_praline-package-2020-05-18.tex};
	\addlegendentry{\bfs}
	
	\addplot+[only marks, color=cb-Set1-1, every mark/.append style={solid, scale=.6}, mark=triangle*]
	table[x=vtcs,y=fd/rand]
	{ncrDA_praline-package-2020-05-18.tex};
	\addlegendentry{\fd}
	
	\end{axis}
	\end{tikzpicture}
	\begin{tikzpicture}
	\begin{axis}[
	width=\textwidth,
	height=.36\textheight,
	legend pos=south east,
	legend cell align={left},
	xmin=0,
	xmax=255,
	ymin=0.5,
	ymax=1.5,
	grid=both,
	ylabel={ratio of bends},
	xlabel={number of vertices}]
	
	\addplot+[only marks, color=cb-Dark2-1,  every mark/.append style={solid, fill=cb-Set1-6, scale=.45}, mark=square*]
	table[x=vtcs,y=bfs/rand]
	{nbpDA_praline-package-2020-05-18.tex};
	\addlegendentry{\bfs}
	
	\addplot+[only marks, color=cb-Set1-1, every mark/.append style={solid, scale=.6}, mark=triangle*]
	table[x=vtcs,y=fd/rand]
	{nbpDA_praline-package-2020-05-18.tex};
	\addlegendentry{\fd}
	
	\end{axis}
	\end{tikzpicture}
	
	\caption{Comparison of the edge-orientation methods \fd and \bfs relative to \rand.
		In each color, each dot represents one of the 380 original plans.}
	\label{fig:oe-plot}
\end{figure}

\begin{figure}[p]
	\centering
	%	\begin{subfigure}{\textwidth}
	\begin{tikzpicture}
	\begin{axis}[
	width=\textwidth,
	height=.54\textheight,
	legend pos=north east,
	legend cell align={left},
	xmin=0,
	xmax=255,
	ymin=0,
	ymax=1.5,
	grid=both,
	ylabel={ratio of crossings},]
	
	\addplot+[only marks, color=cb-Dark2-1,  every mark/.append style={solid, fill=cb-Set1-6, scale=.45}, mark=square*]
	table[x=vtcs,y=bfs/rand]
	{ncrDA_generated_2021-08-06_17-27-03.tex};
	\addlegendentry{\bfs}
	
	\addplot+[only marks, color=cb-Set1-1, every mark/.append style={solid, scale=0.6}, mark=triangle*]
	table[x=vtcs,y=fd/rand]
	{ncrDA_generated_2021-08-06_17-27-03.tex};
	\addlegendentry{\fd}
	\end{axis}
	\end{tikzpicture}
	\begin{tikzpicture}
	\begin{axis}[
	width=\textwidth,
	height=.36\textheight,
	legend pos=south east,
	legend cell align={left},
	xmin=0,
	xmax=255,
	ymin=0.5,
	ymax=1.5,
	grid=both,
	ylabel={ratio of bends},
	xlabel={number of vertices}]
	
	\addplot+[only marks, color=cb-Dark2-1,  every mark/.append style={solid, fill=cb-Set1-6, scale=.45}, mark=square*]
	table[x=vtcs,y=bfs/rand]
	{nbpDA_generated_2021-08-06_17-27-03.tex};
	\addlegendentry{\bfs}
	
	\addplot+[only marks, color=cb-Set1-1, every mark/.append style={solid, scale=0.6}, mark=triangle*]
	table[x=vtcs,y=fd/rand]
	{nbpDA_generated_2021-08-06_17-27-03.tex};
	\addlegendentry{\fd}
	\end{axis}
	\end{tikzpicture}
	
	\caption{Comparison of the edge-orientation methods \fd and \bfs relative to \rand.
		In each color, each dot represents one of the 1140 generated plans.}
	\label{fig:oe-plot-generated}
\end{figure}
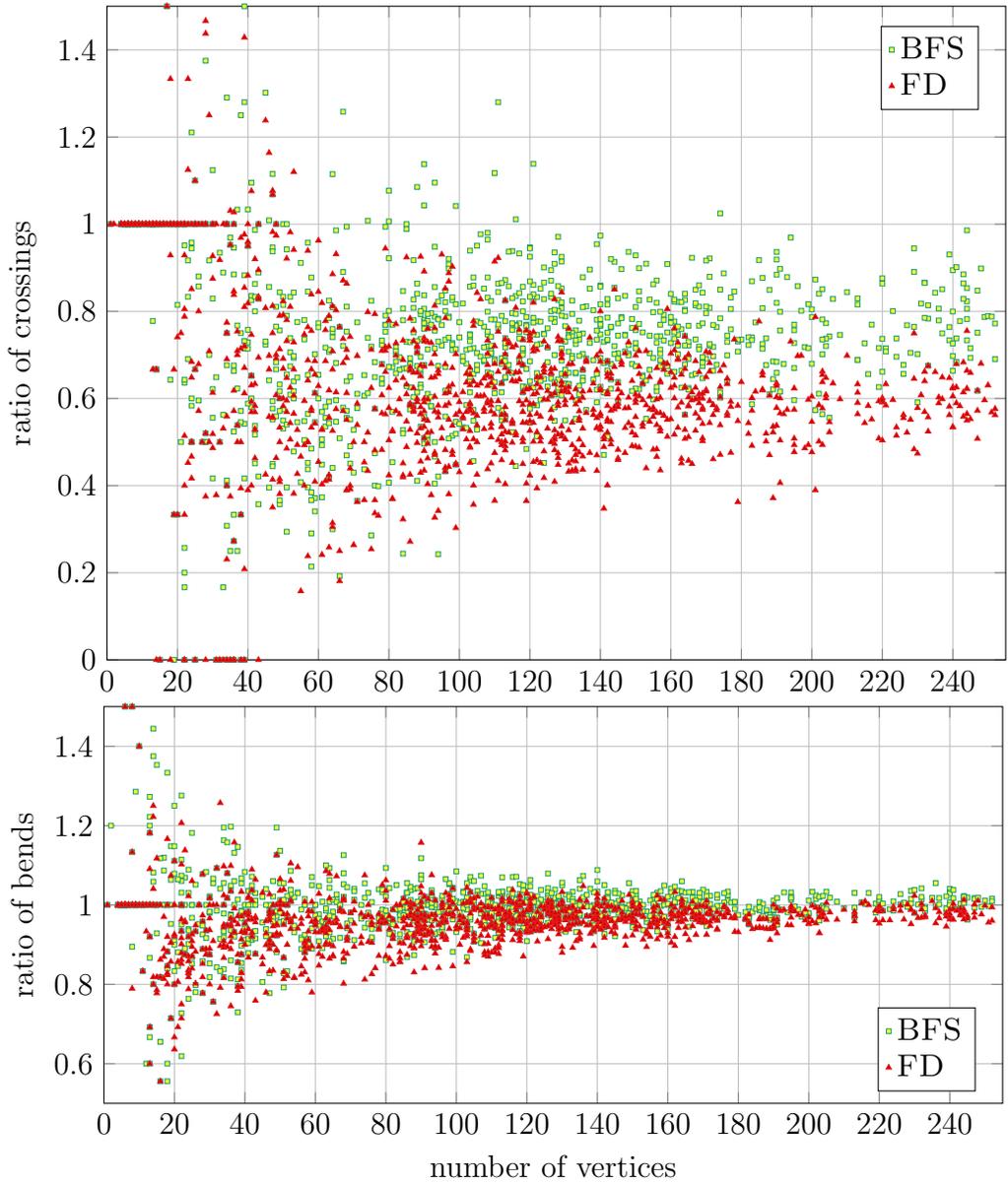

For each graph and each criterion, we took for each method the best of the ten results
and normalized by the best value of \rand.
The means ($\mu$) of these values are listed in Table~\ref{table:results-orienting}.
The winner percentage $\beta$ measures how often a specific method
achieved the best objective value (usually the smallest, but for the
aspect ratio (w:h) the one closest to~1).  Ties are not broken, so
over the three methods, the $\beta$-values add up to more than~100.
We relate the normalized values of~\ncr and~\nbp to the number of
vertices; see Fig.~\ref{fig:oe-plot} for the original plans and
Fig.~\ref{fig:oe-plot-generated} for the generated plans.

\subsubsection{Crossing Reduction}
We used the same settings as when we compared the methods for
orienting the edges, but here we exclusively used \fd for orienting
the edges.  We compared the methods \vertices, \mixed, and \ports,
and the methods \pseudobc, \oppositebc, and \relpos
each with a single run of the crossing reduction phase.  \kieler
joined the comparison as the base line method to which we relate our results.

The variant \kieler uses instead of our algorithm
the algorithm ElkLayered in eclipse.elk (formerly known as: KLayered in KIELER)~\cite{Schulze2020}.
As our algorithm, ElkLayered does Sugiyama-based layered drawing using ports at vertices.
ElkLayered, however, expects a directed graph as input and its port
constraints are less powerful.  ElkLayered offers
free placement of the ports around a vertex, fixed side at a vertex,
fixed order around a vertex, and fixed position at a vertex.
After orienting the given undirected graph,
we used this algorithm as a black box
when we set the port constraints to the most flexible value for each vertex.
So, for vertices having multiple port groups or port pairings, we set the order of ports to be fixed,
while we allow free port placement for all other vertices.
As both algorithms expect different input, use different subroutines and ElkLayered uses more additional steps for producing aesthetic drawings,
this comparison should be treated with caution.

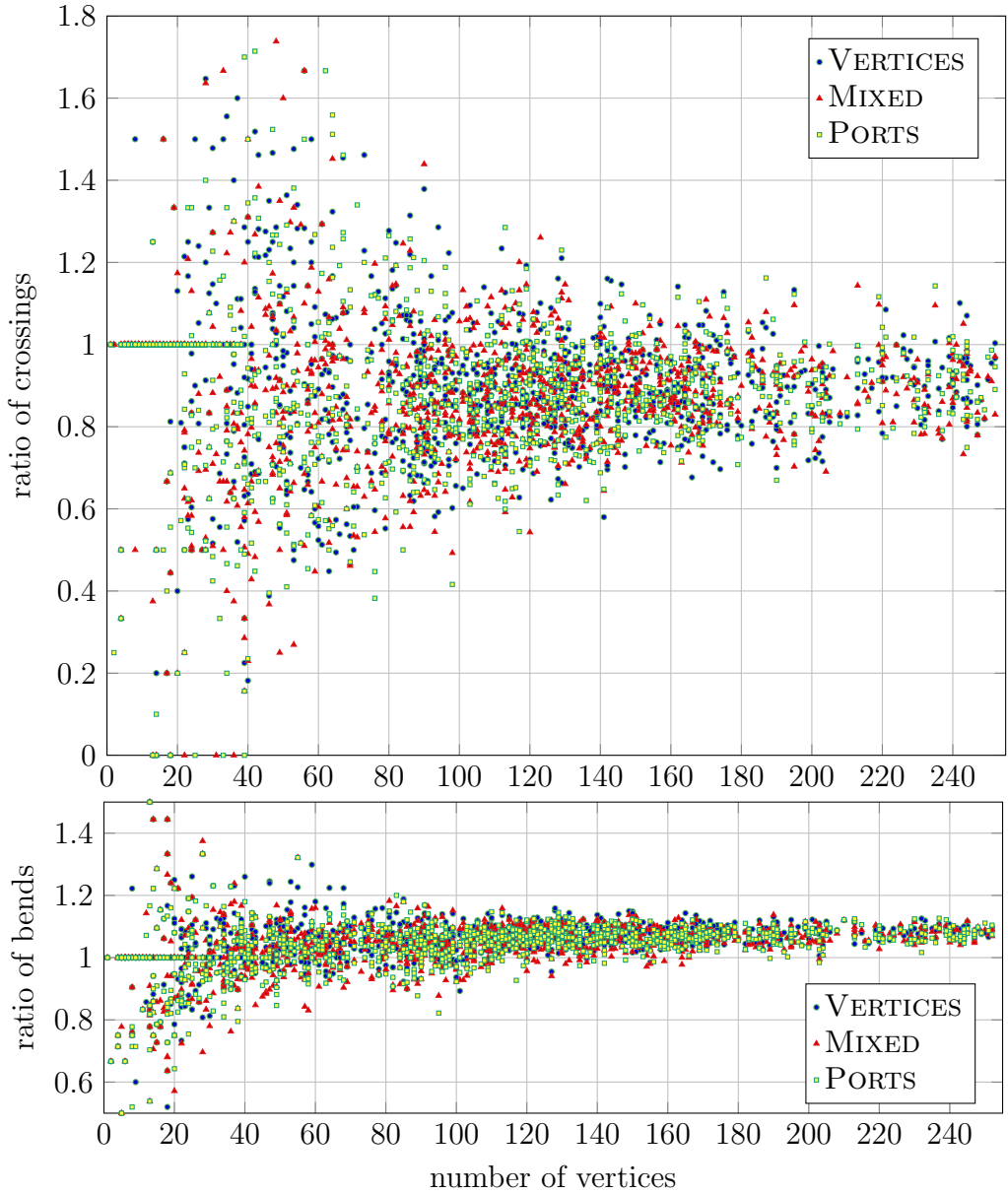
\begin{figure}[p]
	\centering
	\begin{tikzpicture}
	\begin{axis}[
	width=\textwidth,
	height=.60\textheight,
	legend pos=north east,
	legend cell align={left},
	xmin=0,
	xmax=255,
	ymin=0,
	ymax=1.8,
	grid=both,
	ylabel={ratio of crossings},]
	
	\addplot+[only marks, color=cb-Set1-3, every mark/.append style={scale=0.5}]
	table[x=vtcs,y=nodes/kieler]
	{ncrCM_generated_2021-08-06_17-27-03.tex};
	\addlegendentry{\vertices}
	
	\addplot+[only marks, color=cb-Set1-1, every mark/.append style={solid, scale=0.6}, mark=triangle*]
	table[x=vtcs,y=mixed/kieler]
	{ncrCM_generated_2021-08-06_17-27-03.tex};
	\addlegendentry{\mixed}
	
	\addplot+[only marks, color=cb-Dark2-1,  every mark/.append style={solid, fill=cb-Set1-6, scale=0.4}, mark=square*]
	table[x=vtcs,y=ports/kieler]
	{ncrCM_generated_2021-08-06_17-27-03.tex};
	\addlegendentry{\ports}
	
	\end{axis}
	\end{tikzpicture}
	\begin{tikzpicture}
	\begin{axis}[
	width=\textwidth,
	height=.30\textheight,
	legend pos=south east,
	legend cell align={left},
	xmin=0,
	xmax=255,
	ymin=0.5,
	ymax=1.5,
	grid=both,
	ylabel={ratio of bends},
	xlabel={number of vertices}]
	
	\addplot+[only marks, color=cb-Set1-3, every mark/.append style={scale=0.5}]
	table[x=vtcs,y=nodes/kieler]
	{nbpCM_generated_2021-08-06_17-27-03.tex};
	\addlegendentry{\vertices}
	
	\addplot+[only marks, color=cb-Set1-1, every mark/.append style={solid, scale=0.6}, mark=triangle*]
	table[x=vtcs,y=mixed/kieler]
	{nbpCM_generated_2021-08-06_17-27-03.tex};
	\addlegendentry{\mixed}
	
	\addplot+[only marks, color=cb-Dark2-1,  every mark/.append style={solid, fill=cb-Set1-6, scale=0.4}, mark=square*]
	table[x=vtcs,y=ports/kieler]
	{nbpCM_generated_2021-08-06_17-27-03.tex};
	\addlegendentry{\ports}
	
	\end{axis}
	\end{tikzpicture}
	
	\caption{Comparison of the three crossing-reduction methods relative to \kieler.
		For handling local sources and sinks, we have used \relpos.
		In each color, each dot represents one
		of the 1140 generated cable plans.}
	\label{fig:cm-plot-generated}
\end{figure}

For our results, see Table~\ref{table:results-cm}
and Figs.~\ref{fig:cm-plot}/\ref{fig:cm-plot-generated}.

\begin{table}[p]
	\caption{Comparison of the methods for crossing reduction.
		The mean $\mu$ is relative to \kieler; the standard deviation is in the
		range $[.1, .7]$;
		$\beta$ is as in Table~\ref{table:results-orienting}.}
	\label{table:results-cm}
	
	\setlength{\tabcolsep}{2pt}
	\centering
	\medskip
	
	\begin{tabular}{@{}r@{~}|rr@{~~}rr@{~~}rr@{~~}r@{\hspace{1.2ex}}r@{\hspace{1ex}}|@{\hspace{1ex}}rr@{~~}rr@{~~}rr@{~~}r@{}r@{}}
		\toprule
		\multicolumn{1}{c}{} 
		& \multicolumn{8}{c}{Original cable plans}
		& \multicolumn{8}{c}{Generated artificial cable plans}
		\\\midrule
		& \multicolumn{2}{c@{~~}}{\textsc{Vtcs}.}
		& \multicolumn{2}{c@{~~}}{\mixed} 
		& \multicolumn{2}{c@{~~}}{\ports} 
		& \multicolumn{2}{c@{\hspace{1ex}}|@{\hspace{1ex}}}{\textsc{Kiel}.}
		& \multicolumn{2}{c@{~~}}{\textsc{Vtcs}.}
		& \multicolumn{2}{c@{~~}}{\mixed} 
		& \multicolumn{2}{c@{~~}}{\ports} 
		& \multicolumn{2}{c}{\textsc{Kiel}.} \\ 
		& $\mu$ & $\beta$ & $\mu$ & $\beta$ & $\mu$ & $\beta$ & $\mu$ & $\beta$ & $\mu$ & $\beta$ & $\mu$ & $\beta$ & $\mu$ & $\beta$ & $\mu$ & $\beta$ \\
		\midrule
		\multicolumn{1}{c}{} &
		\multicolumn{16}{c}{\pseudobc} \\
		\midrule
		\ncr & 1.57 & 12 & 1.54 & 15 & 1.53 & 16 & 1 & \textbf{83}
		& 1.50 & 11 & 1.52 & 11 & 1.51 & 12 & 1 & \textbf{94} \\
		\nbp & 1.05 & 12 & 1.03 & 25 & 1.03 & 19 & 1 & \textbf{64}
		& 1.06 & 11 & 1.05 & 16 & 1.04 & 17 & 1 & \textbf{79} \\
		width & 1.06 & 17 & 1.06 & 17 & 1.05 & 16 & 1 & \textbf{54}
		& 1.12 & 13 & 1.13 & 11 & 1.12 & 13 & 1 & \textbf{69} \\
		height & 1.36 & 6 & 1.36 & 4 & 1.36 & 5 & 1 & \textbf{91}
		& 1.42 & 1 & 1.43 & 1 & 1.42 & 1 & 1 & \textbf{98} \\
		area & 1.43 & 6 & 1.42 & 7 & 1.42 & 6 & 1 & \textbf{85}
		& 1.60 & 2 & 1.61 & 2 & 1.61 & 2 & 1 & \textbf{97} \\
		w:h & ~.91 & \textbf{30} & ~.91 & 27 & ~.91 & 27 & 1 & 18
		& ~.91 & \textbf{29} & ~.91 & 28 & ~.91 & 29 & 1 & 16 \\
		time & 1.09 & 50 & 1.25 & 14 & 1.31 & 9 & 1 & \textbf{52}
		& 1.45 & 13 & 1.80 & 8 & 1.95 & 7 & 1 & \textbf{92} \\
		\midrule
		\multicolumn{1}{c}{} &
		\multicolumn{16}{c}{\oppositebc} \\
		\midrule
		\ncr & 1.07 & 35 & 1.11 & 22 & 1.03 & \textbf{38} & 1 & 32
		& 1.12 & 36 & 1.22 & 17 & 1.15 & 28 & 1 & \textbf{45} \\
		\nbp & 1.04 & 12 & 1.02 & 26 & 1.03 & 22 & 1 & \textbf{61}
		& 1.05 & 12 & 1.04 & 18 & 1.04 & 17 & 1 & \textbf{75} \\
		width & 1.13 & 17 & 1.13 & 14 & 1.14 & 15 & 1 & \textbf{59}
		& 1.23 & 6 & 1.25 & 6 & 1.24 & 6 & 1 & \textbf{89} \\
		height & 1.31 & 6 & 1.32 & 6 & 1.31 & 4 & 1 & \textbf{90}
		& 1.38 & 1 & 1.38 & 1 & 1.38 & 1 & 1 & \textbf{98} \\
		area & 1.48 & 6 & 1.50 & 7 & 1.49 & 4 & 1 & \textbf{87}
		& 1.72 & 2 & 1.73 & 2 & 1.72 & 2 & 1 & \textbf{97} \\
		w:h & ~.93 & \textbf{33} & ~.93 & 23 & ~.93 & 27 & 1 & 19
		& ~.95 & 29 & ~.95 & 29 & ~.95 & \textbf{29} & 1 & 15 \\
		time & 1.61 & 23 & 1.96 & 11 & 1.81 & 13 & 1 & \textbf{75}
		& 2.30 & 10 & 2.77 & 7 & 2.83 & 6 & 1 & \textbf{94} \\
		\midrule
		\multicolumn{1}{c}{} &
		\multicolumn{16}{c}{\relpos} \\
		\midrule
		\ncr & ~.82 & 23 & ~.72 & 47 & ~.70 & \textbf{54} & 1 & 9
		& ~.92 & 35 & ~.90 & 41 & ~.89 & \textbf{42} & 1 & 16 \\
		\nbp & 1.04 & 13 & 1.03 & 20 & 1.02 & 26 & 1 & \textbf{60}
		& 1.06 & 10 & 1.04 & 18 & 1.04 & 18 & 1 & \textbf{75} \\
		width & 1.11 & 13 & 1.08 & 19 & 1.08 & 17 & 1 & \textbf{54}
		& 1.21 & 6 & 1.20 & 8 & 1.21 & 6 & 1 & \textbf{86} \\
		height & 1.29 & 5 & 1.29 & 5 & 1.29 & 5 & 1 & \textbf{91}
		& 1.38 & 1 & 1.36 & 1 & 1.37 & 1 & 1 & \textbf{99} \\
		area & 1.43 & 7 & 1.39 & 9 & 1.40 & 7 & 1 & \textbf{81}
		& 1.67 & 2 & 1.65 & 2 & 1.66 & 2 & 1 & \textbf{97} \\
		w:h & ~.93 & 27 & ~.93 & 25 & ~.92 & \textbf{31} & 1 & 20
		& ~.94 & 26 & ~.94 & \textbf{31} & ~.94 & 30 & 1 & 15 \\
		time & 1.07 & 48 & 1.20 & 12 & 1.24 & 11 & 1 & \textbf{51}
		& 1.43 & 13 & 1.76 & 9 & 1.86 & 8 & 1 & \textbf{92} \\
		\bottomrule
	\end{tabular}
\end{table}
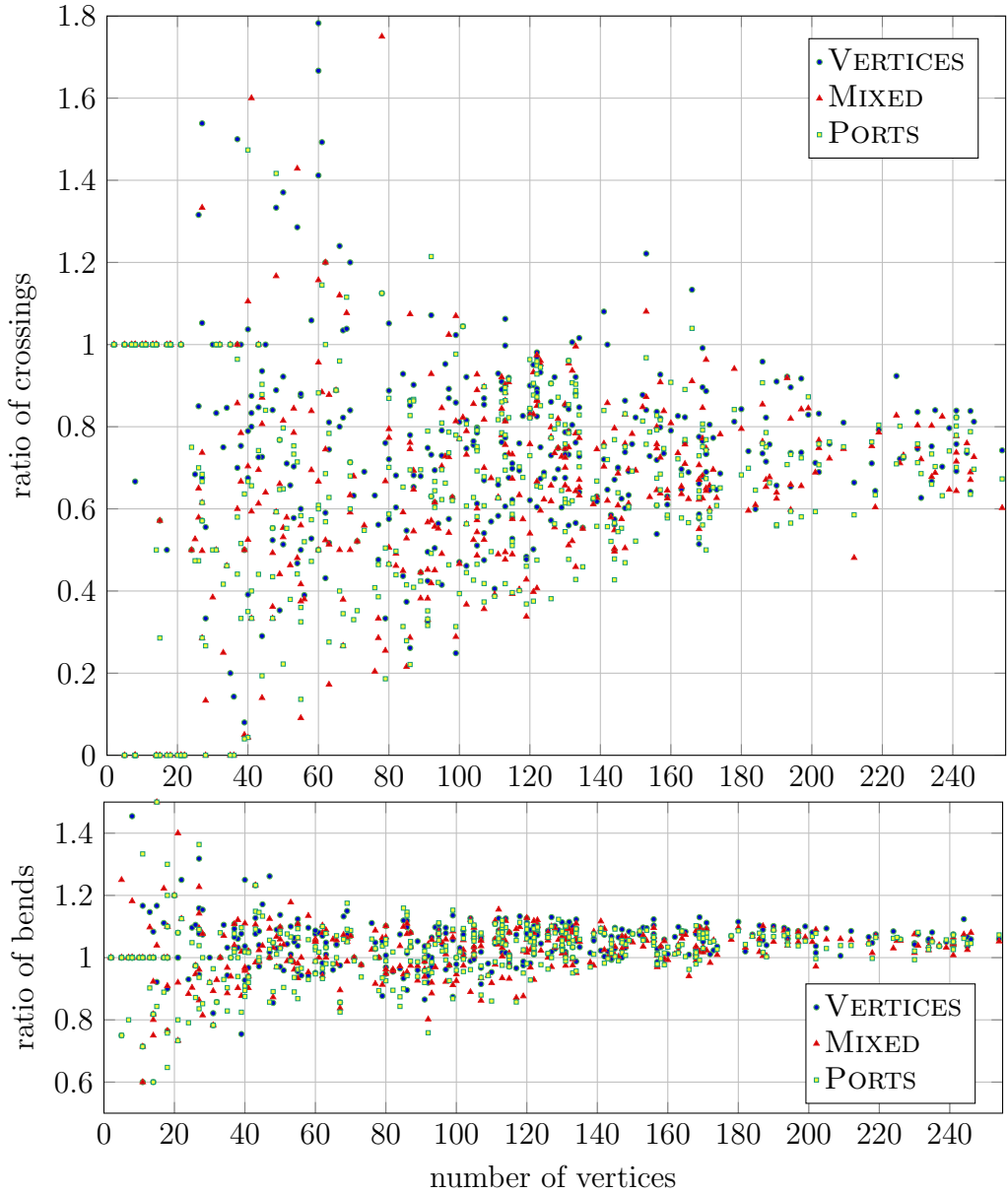
\begin{figure}[p]
	\centering
	%	\begin{subfigure}{\textwidth}
	\begin{tikzpicture}
	\begin{axis}[
	width=\textwidth,
	height=.60\textheight,
	legend pos=north east,
	legend cell align={left},
	xmin=0,
	xmax=255,
	ymin=0,
	ymax=1.8,
	grid=both,
	ylabel={ratio of crossings},]
	
	\addplot+[only marks, color=cb-Set1-3, every mark/.append style={scale=.5}]
	table[x=vtcs,y=nodes/kieler]
	{ncrCM_praline-package-2020-05-18.tex};
	\addlegendentry{\vertices}
	
	\addplot+[only marks, color=cb-Set1-1, every mark/.append style={solid, scale=.6}, mark=triangle*]
	table[x=vtcs,y=mixed/kieler]
	{ncrCM_praline-package-2020-05-18.tex};
	\addlegendentry{\mixed}
	
	\addplot+[only marks, color=cb-Dark2-1,  every mark/.append style={solid, fill=cb-Set1-6, scale=0.4}, mark=square*]
	table[x=vtcs,y=ports/kieler]
	{ncrCM_praline-package-2020-05-18.tex};
	\addlegendentry{\ports}
	
	\end{axis}
	\end{tikzpicture}
	%	\end{subfigure}
	%	\smallskip
	%	\begin{subfigure}{\textwidth}
	\begin{tikzpicture}
	\begin{axis}[
	width=\textwidth,
	height=.30\textheight,
	legend pos=south east,
	legend cell align={left},
	xmin=0,
	xmax=255,
	ymin=0.5,
	ymax=1.5,
	grid=both,
	ylabel={ratio of bends},
	xlabel={number of vertices}]
	
	\addplot+[only marks, color=cb-Set1-3, every mark/.append style={scale=.5}]
	table[x=vtcs,y=nodes/kieler]
	{nbpCM_praline-package-2020-05-18.tex};
	\addlegendentry{\vertices}
	
	\addplot+[only marks, color=cb-Set1-1, every mark/.append style={solid, scale=.6}, mark=triangle*]
	table[x=vtcs,y=mixed/kieler]
	{nbpCM_praline-package-2020-05-18.tex};
	\addlegendentry{\mixed}
	
	\addplot+[only marks, color=cb-Dark2-1,  every mark/.append style={solid, fill=cb-Set1-6, scale=.4}, mark=square*]
	table[x=vtcs,y=ports/kieler]
	{nbpCM_praline-package-2020-05-18.tex};
	\addlegendentry{\ports}
	\end{axis}
	\end{tikzpicture}
	%	\end{subfigure}
	
	\caption{Comparison of the three crossing-reduction methods
		relative to \kieler.
		For handling local sources and sinks, we used \relpos.
		In each color, each dot represents one of the 380 original cable plans.}
	\label{fig:cm-plot}
\end{figure}

\section{Discussion and Conclusion}

In this section, we discuss the findings of our experiments
in regards of the following aspects.

\subsection{Methods for Orienting Undirected Edges}
\fd almost always yields orientations of the undirected graphs that
lead to drawings with fewer crossings than the orientations obtained
from \bfs and \rand.
The gap between \fd and \bfs is minor,
whereas the gap between both \fd and \bfs to \rand is large.
Regarding the bend points, there is a rather negligible advantage
for \fd and \bfs.
Comparing the drawing area, \fd and \bfs are
similar, but \fd achieves a better aspect ratio.
Although \rand performs
rather poorly for most criteria, it often uses the smallest drawing area.
The savings in the total area by \rand can be attributed
almost exclusively to a small height, which corresponds to fewer layers.

The layer assignment procedure uses more layers if we have longer
paths of directed edges.  \fd rather straightens a path between two
(distant) vertices requiring then more layers, while \rand rather
orients some of the edges of this path up and some down, yielding
shorter chains of directed edges.  So, \rand has more vertices per
layer, which also explains the worse width and aspect ratio.
We suspect that this large width
might partially be explained by the use of the algorithm of
Brandes and K\"opf~\cite{Brandes2001} in the coordinate assignment
phase.  In this phase, many edges are
drawn vertically.  After the crossing minimization phase, we would
expect that the vertices on the layers come close to the initial
non-layered drawing of \fd having short edges.  When the edges between
each two layers are longer for \rand, straightening them to a vertical
line segment pushes vertices on the upper layer further apart from
vertices on the lower layer.

Comparing the running times of the three variants, we note that using
\fd is about 10--20\,\% slower and using \bfs is about 10-20\,\%
faster than using \rand.
We remark that these percentages refer to the running time of the
whole algorithm, not just to the edge orienting phase.
This explains why \rand is not necessarily the fastest variant;
% The execution time of the edge orienting phase using \bfs
% might be greater than the execution time of this phase using \rand,
% however the next steps might take longer when using \rand
% because the resulting intermediate graphs are different.
e.g., if \rand produces many dummy vertices and wider layers,
the crossing reduction phase may take longer.

Summing up, we remark that it is worth using
a more sophisticated method (\fd or \bfs) for
orienting the undirected edges than just using a
random assignment (\rand).
The choice between \fd and \bfs depends on the user's preferences.
\fd tends to produce fewer crossings and a more balanced aspect
ratio.  \bfs, in contrast, is (slightly) faster
and conceptually simpler to understand and to implement.
As our main goal is obtaining visually pleasant drawings,
we recommend using \fd for orienting edges if
a (fast) force-directed graph drawing algorithm is available.

\subsection{Methods for Crossing Reduction}
We first consider the method for handling local sources and sinks in
the layer-sweep algorithm.
Then we analyze the methods for treating ports and vertices
and compare them to \kieler.

\paragraph{Methods for Handling Local Sources and Sinks}
Regarding the number of edge crossings, the rather simple approach
\relpos outperforms \pseudobc and \oppositebc by far.
The second-best method is clearly \oppositebc, whereas
\pseudobc performs rather poorly.
Regarding the number of bends and the drawing area, all three
approaches behave quite similarly.
\relpos and \pseudobc are about 50--70\,\% faster than \oppositebc,
with a slight advantage for \relpos.

In our experiments, \relpos turned out to clearly be the best method
or at least as good as the others, both in terms of simplicity and in
terms of the criteria we measured.  Therefore, we recommend \relpos,
and in all remaining experiments, we use \relpos.

\paragraph{Methods for Treating Ports and Vertices}
In terms of number of edge crossings, the methods \ports and \mixed
achieve similar results; both clearly beat the method \vertices.
This is in line with our expectation that incorporating distinct port orderings
during the whole crossing reduction procedure helps to
avoid edge crossings, which crucially depend on the precise order of ports.
However, incorporating all ports (\ports) instead of only ports at vertices
with port pairings (\mixed) does not seem to provide much of an additional benefit.
(Recall that \mixed is not a generalization of \ports, but rather a
generalization of \vertices as the whole accounting is vertex-based
instead of port-based.)

Regarding the number of bends and the drawing area, all variants
perform similarly well.
As expected, using \vertices is faster than
using \mixed, which in turn is faster than \ports.
Using \ports, the total running time increases by about 10--40\,\%
with respect to \vertices.

Since we deem a small number of crossings the most important
quality measure, we recommend using \mixed or \ports,
which we consider both equally well suited for our application.

\subsection{Comparison to \kieler}
Regarding the number of edge crossings, our new methods outperform the existing
algorithm, which has not been designed for these specific port constraints.
For the original cable plans, \mixed and \ports use about
30\,\% fewer crossings and \vertices still achieves about 20\,\% fewer crossings.

The number of bends is about the same for all variants of our algorithm
and \kieler---we use in average at most 6\,\% more bends.
We remark that this highly depends on the width of our vertex rectangles.
Remember that we have adjusted the algorithm of Brandes and K{\"o}pf~\cite{Brandes2001}
to handle ports instead of vertices and to limit the distance of ports within
the same vertex.
Allowing an arbitrary placement also for ports of the same vertex leads to fewer bends,
but also produces drawings with overwide vertices.
In an earlier version of our algorithm~\cite{Walter2020a},
we did not limit the distance between ports within the same vertex.
Additionally, our implementations differed in some other minor aspects.
This resulted in our variants using much fewer bends than \kieler.
Now we have made a design choice to avoid large gaps between ports
within a vertex.  (Recall that we break vertical alignments if gaps
are larger than 16 times the minimum port distance.)
We observed that due to this choice (a)~the drawings are sufficiently
compact and (b)~vertex rectangles have an appealing aspect ratio.
Moreover, we use roughly as many bends as \kieler does.

The drawings generated by our new algorithm use an about
40\,\% (original plans) and 65\,\% (artificial plans) larger area 
than the ones generated by \kieler.
The main difference comes from a greater height,
which we get from more horizontal lines being used for the orthogonal edge routing
and for integrating the intermediate layers that we use for
turning dummy vertices.
However, as the drawings generated by \kieler tend to be wider than high,
using a greater height leads to a better aspect ratio for our variants
(better in the sense of being closer to~1, i.e., the bounding box being
more square-like).

Also with respect to the running time, \kieler produces its drawing in average
a little faster than our algorithm.
On the original plans, our variants need in average almost the same time (\vertices)
or about 25\,\% more time (\ports).
This gap is larger for the generated artificial cable plans,
but still seems to be in the range from a factor of 1 to a factor of 2 compared to \kieler.
In total numbers, \vertices, \mixed, \ports, and \kieler needed in
average for the original plans 142\,ms, 166\,ms, 173\,ms, and 127\,ms,
respectively.  The maximum running time that we measured occurred in a
cable plan with 354 vertices.  It took 1.1\,s, 1.3\,s, 1.6\,s, and
0.6\,s, respectively.

In conclusion, we can say that there is no algorithm being superior in all considered aspects.
Cognitive studies, however, have shown that a small number of crossings
highly influences the readability of a graph drawing for a human user~\cite{Purchase2002,Ware2002}.
Our industry partners gave us similar feedback when working with these cable plans.
Therefore, we consider reducing the number of crossings
by more than a fourth and almost a third to be more important than a slightly smaller drawing area
(which is likely to be less readable) and a slightly faster running
time (which has to be done only once).
%\todo{AW: Was meinst du hier mit ``in advance''?
%JZ: ich meine, dass es nur einmal im Vorfeld berechnet werden muss.
%Moeglicherweise ist es misseverstaendlich;
%ich habe es umformuliert in ``muss nur einmal gemacht werden''.}
Therefore, we recommend using our new algorithm with the variants
\ports or \mixed in combination with \relpos when working with
generalized port constraints and---more specifically---when
working with cable or circuit plans that are somehow similar to the
ones in our experiments.

We remark that the application settings that \kieler is designed for is not the same
as for our algorithm, which limits the meaningfulness of this comparison.
Moreover \kieler uses more intermediate steps and post-processing steps,
e.g., for compactification, which partially explains the smaller drawing area.
\kieler also has more additional functionalities and is the overall
more mature and established library.
\kieler also provided an excellent starting point for our research and
helped us to quickly generate some initial layered cable and circuit
plans for our industrial partners.

\subsection{Generating Pseudo Cable Plans}
We concede that the artificial plans that we generated are not perfect
as they behave somewhat differently from the original plans for
certain criteria.  For instance, for the artificial plans the relative
advantage of \ports and \mixed compared to \vertices vanishes.
Also our variants perform worse compared to \kieler with respect to
the number of edge crossings, drawing area, and running time.
%The plots in Figs.~\ref{fig:oe-plot-generated}
%and~\ref{fig:cm-plot-generated} look more unstructured
%than those in Figs.~\ref{fig:oe-plot} and~\ref{fig:cm-plot}.
Nevertheless, the obfuscation allowed us to make somewhat realistic
cable plans publicly available, so that others can
validate our experiments in the future.

\subsection{Open Problems}
Our generation procedure may also serve as an entry point for more research
in generating pseudo data from original data.
This approach can be applied in many domains (and has most probably been
applied, in domains we are not aware of).
Finding such connections and formalizing the theory behind our
obfuscation procedure would be interesting.

We are currently in the process of integrating our algorithm into the
software of our industrial partner.
We hope to see whether the statistical
improvement of our algorithm actually yields advantages in practice.
We also hope for practically relevant feedback and problems,
which we can theoretically formalize and integrate in our model and algorithm.

We have not yet investigated much the usual tuning of parameters,
e.g., the number of repetitions for the crossing reduction phase (currently $r = 1$)
or more repetitions of the whole procedure.
Beside minor tuning, our algorithm still leaves room for more radical
improvements in many spots.
This regards mainly the crossing reduction phase, the node/port placement phase,
and the edge routing phase.

It turned out that edge routing gives rise to a cute combinatorial problem
(see Section~\ref{sub:edge-routing}), which we have not yet solved completely.
Let us recapitulate the problem here.
We are given a set line segments whose (distinct) endpoints
lie on two parallel lines $L_i$ and $L_{i+1}$, i.e., a pair of layers in our drawing.
Each line segment must be re-drawn as an orthogonal polyline
consisting of a vertical, a horizontal, and a second vertical piece.
The horizontal pieces must be placed onto parallel horizontal lines
$\ell_1$, $\ell_2$, $\dots$ lying in between~$L_i$ and~$L_{i+1}$,
such that no horizontal pieces overlap and each two polylines intersect at most once.
The objective is to minimize the number of horizontal lines being occupied
by horizontal pieces.

Recently, Br{\"u}ckner~\cite{b-ozfppg-BT21}
and Mittelst{\"a}dt~\cite{Mittelstaedt2022} have examined
this problem a little deeper.
The set of lines segments induces a conflict graph~$C$ with
both directed and undirected edges.
An undirected edge means the horizontal pieces of the
corresponding two polylines must be placed onto distinct lines,
and a directed edge additionally states which horizontal piece
must be above which other horizontal piece to avoid double intersections
between polylines.
Br{\"u}ckner showed that the transitive closure of~$C$ is weakly chordal.
Orienting the undirected edges of~$C$ while minimizing the length
of a longest directed path would provide an assignment of
horizontal pieces to lines, which uses a minimum number of lines.
Mittelst{\"a}dt showed that our
greedy approach is optimal for only left-going (right-going) edges
and hence a 2-approximation for the combination of both.
It would be interesting to see an optimal polynomial-time algorithm
or to show that the problem is NP-hard and to give a better approximation
algorithm if possible.

We are also interested in more domains where we can apply
the concept of layered graph drawing with generalized port constraints---both
for directed and undirected graphs.
Beside cable plans, applications may include circuit plans, IT network
plans, UML diagrams, data-flow networks, knowledge graphs, and many more.

\bibliographystyle{elsarticle-num}
\bibliography{abbrv,layered-graph-drawing}

\begin{thebibliography}{10}
\expandafter\ifx\csname url\endcsname\relax
  \def\url#1{\texttt{#1}}\fi
\expandafter\ifx\csname urlprefix\endcsname\relax\def\urlprefix{URL }\fi
\expandafter\ifx\csname href\endcsname\relax
  \def\href#1#2{#2} \def\path#1{#1}\fi

\bibitem{Sugiyama1981}
K.~Sugiyama, S.~Tagawa, M.~Toda, Methods for visual understanding of
  hierarchical system structures, IEEE Trans. Syst. Man Cybern. 11~(2) (1981)
  109--125.
\newblock \href {https://doi.org/10.1109/TSMC.1981.4308636}
  {\path{doi:10.1109/TSMC.1981.4308636}}.

\bibitem{Biedl2000}
T.~C. Biedl, B.~Madden, I.~G. Tollis, The three-phase method: {A} unified
  approach to orthogonal graph drawing, Int. J. Comput. Geom. Appl. 10~(6)
  (2000) 553--580.
\newblock \href {https://doi.org/10.1142/S0218195900000310}
  {\path{doi:10.1142/S0218195900000310}}.

\bibitem{ssh-dlgpc-JVLC14}
C.~D. Schulze, M.~Sp{\"{o}}nemann, R.~von Hanxleden, Drawing layered graphs
  with port constraints, J. Vis. Lang. Comput. 25~(2) (2014) 89--106.
\newblock \href {https://doi.org/10.1016/j.jvlc.2013.11.005}
  {\path{doi:10.1016/j.jvlc.2013.11.005}}.

\bibitem{Okka2021}
A.~Okka, U.~Dogrusoz, H.~Balci, {CoSEP:} {A} compound spring embedder layout
  algorithm with support for ports, Inform. Vis. 20~(2--3) (2021) 151--169.
\newblock \href {https://doi.org/10.1177/14738716211028136}
  {\path{doi:10.1177/14738716211028136}}.

\bibitem{ew-dg2l-TCS94}
P.~Eades, S.~Whitesides, Drawing graphs in two layers, Theor. Comput. Sci.
  131~(2) (1994) 361--374.
\newblock \href {https://doi.org/10.1016/0304-3975(94)90179-1}
  {\path{doi:10.1016/0304-3975(94)90179-1}}.

\bibitem{Brandes2001}
U.~Brandes, B.~K{\"{o}}pf, Fast and simple horizontal coordinate assignment,
  in: P.~Mutzel, M.~J{\"{u}}nger, S.~Leipert (Eds.), Proc. 9th Int. Symp. Graph
  Drawing (GD'01), Vol. 2265 of LNCS, Springer, 2002, pp. 31--44.
\newblock \href {https://doi.org/10.1007/3-540-45848-4\_3}
  {\path{doi:10.1007/3-540-45848-4\_3}}.

\bibitem{Brandes2020}
U.~Brandes, J.~Walter, J.~Zink, \href{http://arxiv.org/abs/2008.01252}{Erratum:
  Fast and simple horizontal coordinate assignment}, CoRR abs/2008.01252
  (2020).
\newline\urlprefix\url{http://arxiv.org/abs/2008.01252}

\bibitem{fr-gdfdp-SPE91}
T.~M.~J. Fruchterman, E.~M. Reingold, Graph drawing by force-directed
  placement, Softw.~-- Pract. \& Exper. 21~(11) (1991) 1129--1164.
\newblock \href {https://doi.org/10.1002/spe.4380211102}
  {\path{doi:10.1002/spe.4380211102}}.

\bibitem{Lipp2016}
F.~Lipp, A.~Wolff, J.~Zink, Faster force-directed graph drawing with the
  well-separated pair decomposition, Algorithms 9~(3) (2016) 53.
\newblock \href {https://doi.org/10.3390/a9030053}
  {\path{doi:10.3390/a9030053}}.

\bibitem{Gansner1993}
E.~R. Gansner, E.~Koutsofios, S.~C. North, K.~Vo, A technique for drawing
  directed graphs, IEEE Trans. Softw. Engineer. 19~(3) (1993) 214--230.
\newblock \href {https://doi.org/10.1109/32.221135}
  {\path{doi:10.1109/32.221135}}.

\bibitem{Walter2020a}
J.~Walter, J.~Zink, J.~Baumeister, A.~Wolff, Layered drawing of undirected
  graphs with generalized port constraints, in: Proc. 28th Int. Symp. Graph
  Drawing \& Network Vis. (GD'20), Vol. 12590 of LNCS, Springer, 2020, pp.
  220--234.
\newblock \href {https://doi.org/10.1007/978-3-030-68766-3\_18}
  {\path{doi:10.1007/978-3-030-68766-3\_18}}.

\bibitem{Mittelstaedt2022}
F.~Mittelst{\"a}dt,
  \href{https://www1.pub.informatik.uni-wuerzburg.de/pub/theses/2022-mittelstaedt-masterarbeit.pdf}{About
  coloring of generalized interval graphs}, master's thesis, Institut f\"ur
  Informatik, Universit\"at W\"urzburg, in German (2022).
\newline\urlprefix\url{https://www1.pub.informatik.uni-wuerzburg.de/pub/theses/2022-mittelstaedt-masterarbeit.pdf}

\bibitem{github-praline}
\href{https://github.com/j-zink-wuerzburg/praline}{{PRALINE} data structure and
  layouting algorithm} (2020).
\newline\urlprefix\url{https://github.com/j-zink-wuerzburg/praline}

\bibitem{github-pseudo-plans}
\href{https://github.com/j-zink-wuerzburg/\\
  pseudo-praline-plan-generation}{{PRALINE} pseudo plans -- algorithm and data
  sets} (2020).
\newline\urlprefix\url{https://github.com/j-zink-wuerzburg/\\
  pseudo-praline-plan-generation}

\bibitem{Schulze2020}
\href{https://www.eclipse.org/elk/}{Eclipse layout kernel ({ELK})} (2020).
\newline\urlprefix\url{https://www.eclipse.org/elk/}

\bibitem{Purchase2002}
H.~C. Purchase, D.~A. Carrington, J.~Allder, Empirical evaluation of
  aesthetics-based graph layout, Empirical Softw. Engin. 7~(3) (2002) 233--255.
\newblock \href {https://doi.org/10.1023/A:1016344215610}
  {\path{doi:10.1023/A:1016344215610}}.

\bibitem{Ware2002}
C.~Ware, H.~C. Purchase, L.~Colpoys, M.~McGill, Cognitive measurements of graph
  aesthetics, Inform. Vis. 1~(2) (2002) 103--110.
\newblock \href {https://doi.org/10.1057/palgrave.ivs.9500013}
  {\path{doi:10.1057/palgrave.ivs.9500013}}.

\bibitem{b-ozfppg-BT21}
L.~Br{\"u}ckner,
  \href{https://www1.pub.informatik.uni-wuerzburg.de/pub/theses/2021-brueckner-bachelor.pdf}{Orthogonal
  drawing as a coloring problem in perfect graphs}, bachelor's thesis, Institut
  f\"ur Informatik, Universit\"at W\"urzburg, in German (2021).
\newline\urlprefix\url{https://www1.pub.informatik.uni-wuerzburg.de/pub/theses/2021-brueckner-bachelor.pdf}

\end{thebibliography}

\clearpage
\appendix

\section{Cable Plan Drawings}
\label{app:drawings}
Next, we provide drawings of six cable plans (three original plans and
three pseudo plans).
For each plan, there is a drawing generated by our algorithm using
\fd, \ports and \relpos, and there is another drawing generated using \kieler.
The drawings have been generated automatically in a run where each plan
has been drawn ten times and the best drawing (with respect to the
number of crossings) has been kept.
Port pairings are indicated by line segments inside a vertex.

\begin{figure}[h]
	\centering
	\includegraphics[width=\textwidth]{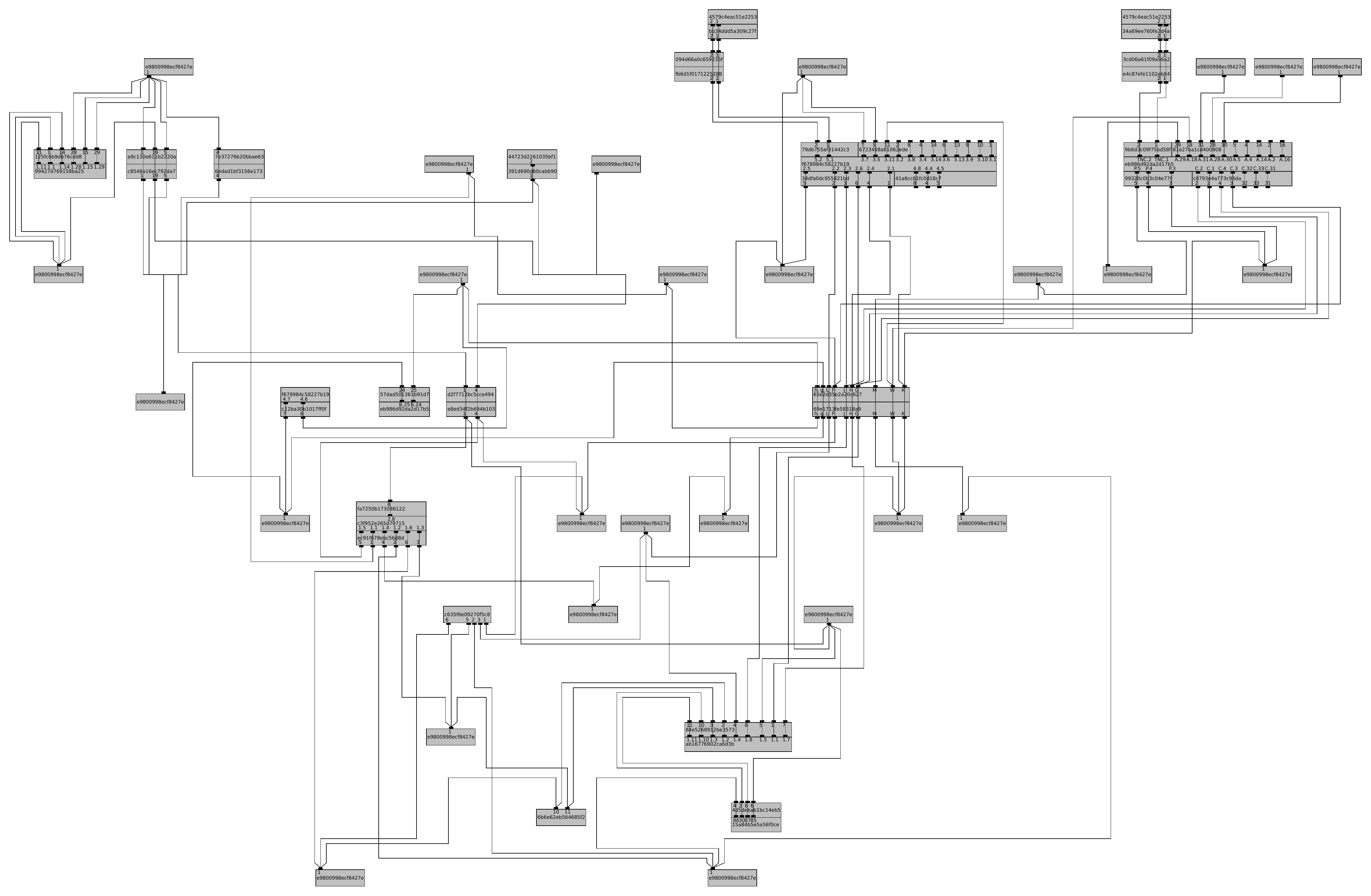}
	\caption{Original cable plan (anonymized) from the large data set with 69 vertices and 104 crossings drawn by our algorithm using \fd, \ports and \relpos.}
	\label{fig:orig-large-ports-79vtcs}
\end{figure}

\begin{figure}[p]
	\centering
	\includegraphics[width=\textwidth]{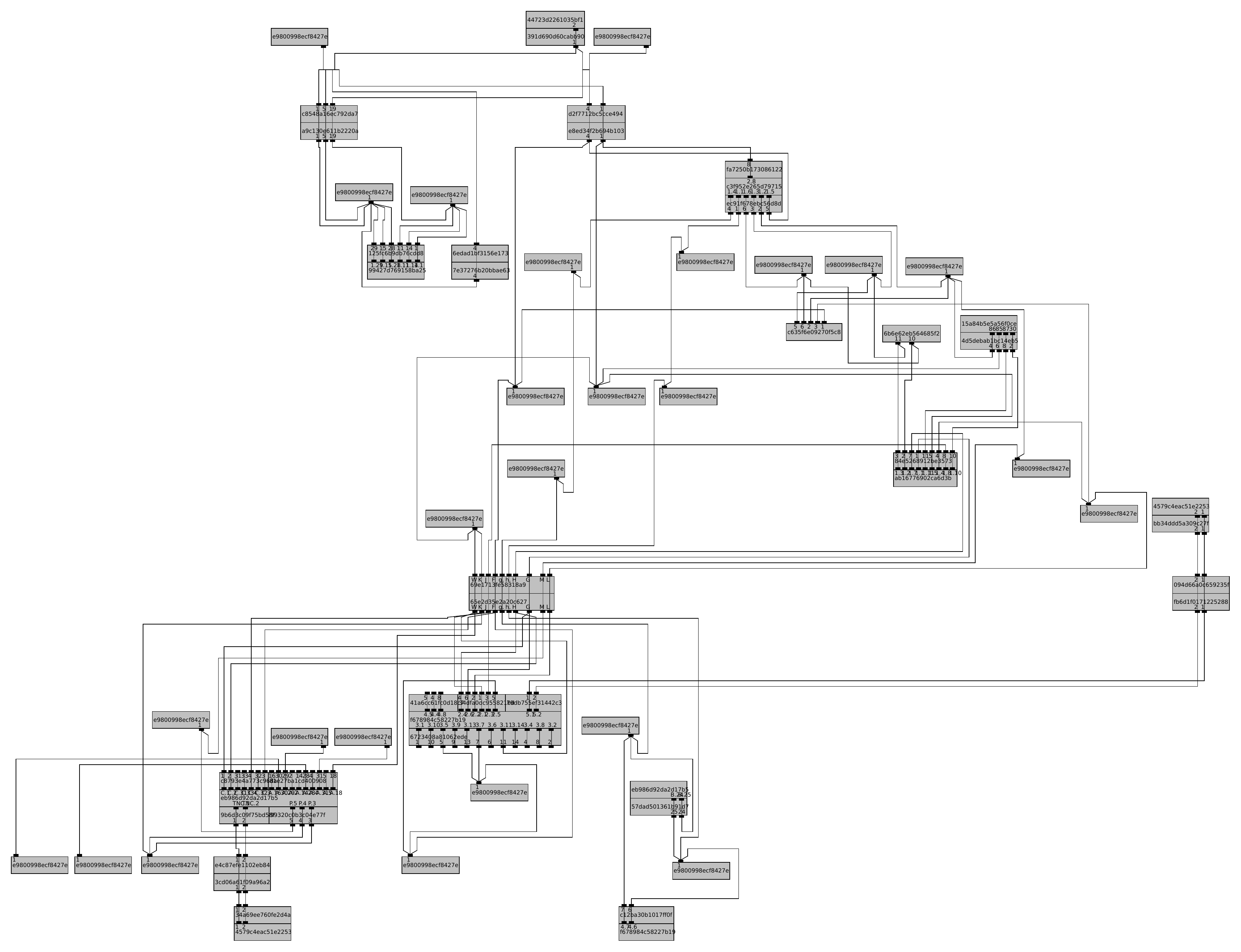}
	\caption{Original cable plan (anonymized) from the large data set with 69 vertices and 162 crossings drawn using \kieler.}
	\label{fig:orig-large-kieler-79vtcs}
\end{figure}

\begin{figure}[p]
	\centering
	\includegraphics[width=\textwidth]{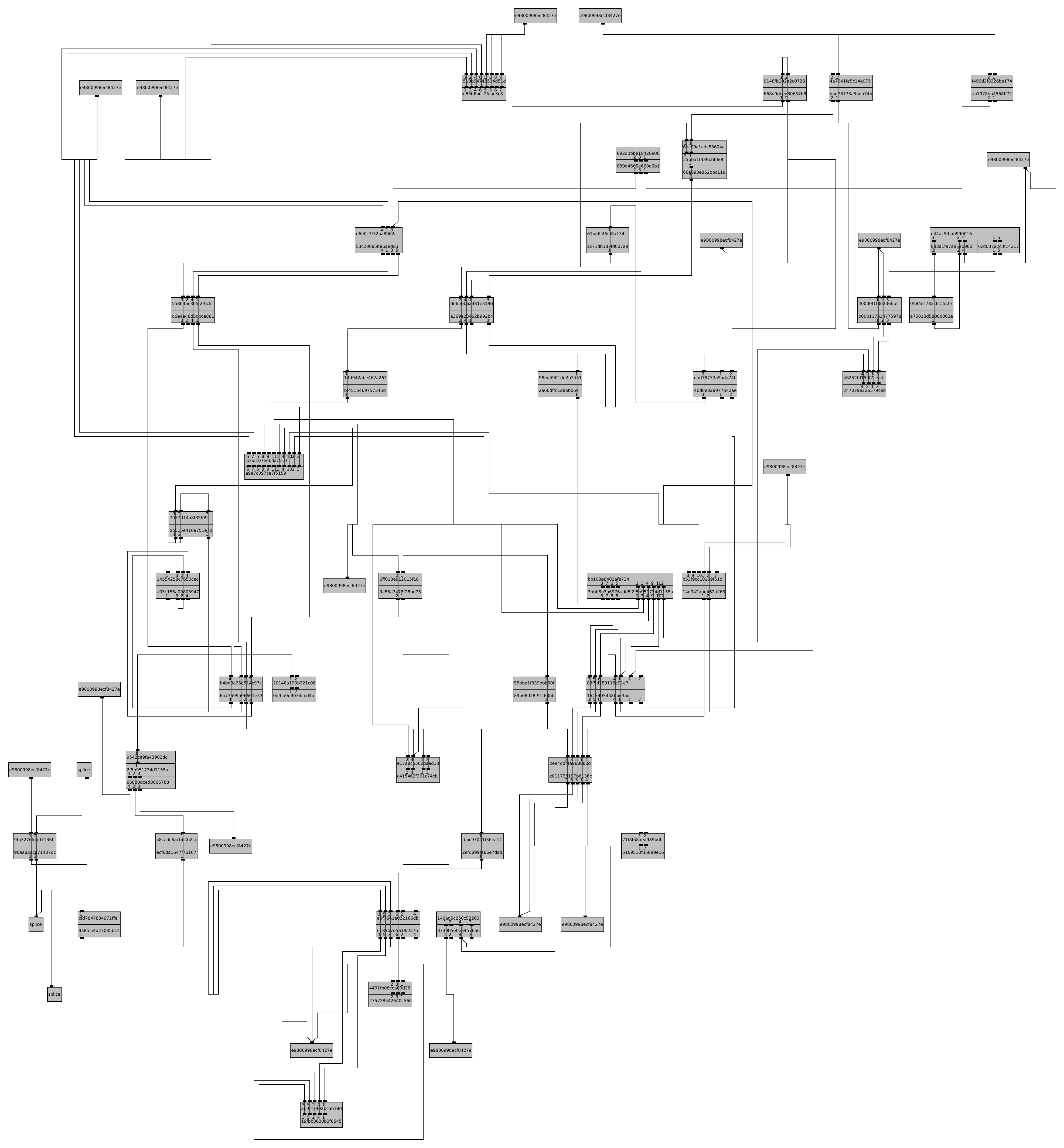}
	\caption{Pseudo cable plan generated from the large data set with 101 vertices and 138 crossings drawn by our algorithm using \fd, \ports and \relpos.}
	\label{fig:pseudo-large-ports-101vtcs}
\end{figure}

\begin{figure}[p]
	\centering
	\includegraphics[width=\textwidth]{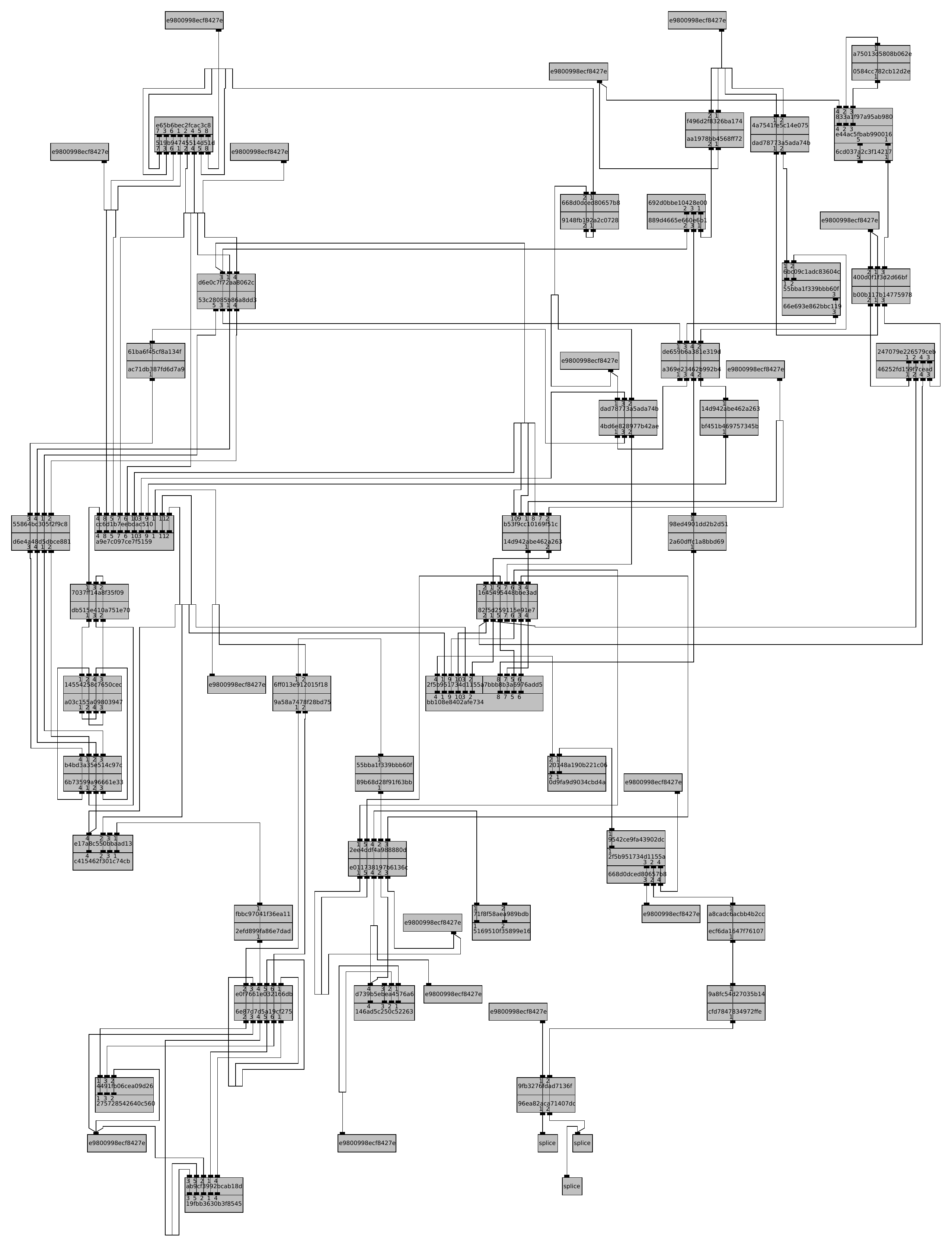}
	\caption{Pseudo cable plan generated from the large data set with 101 vertices and 149 crossings drawn using \kieler.}
	\label{fig:pseudo-large-kieler-101vtcs}
\end{figure}

\begin{figure}[p]
	\centering
	\includegraphics[height=.9\textheight]{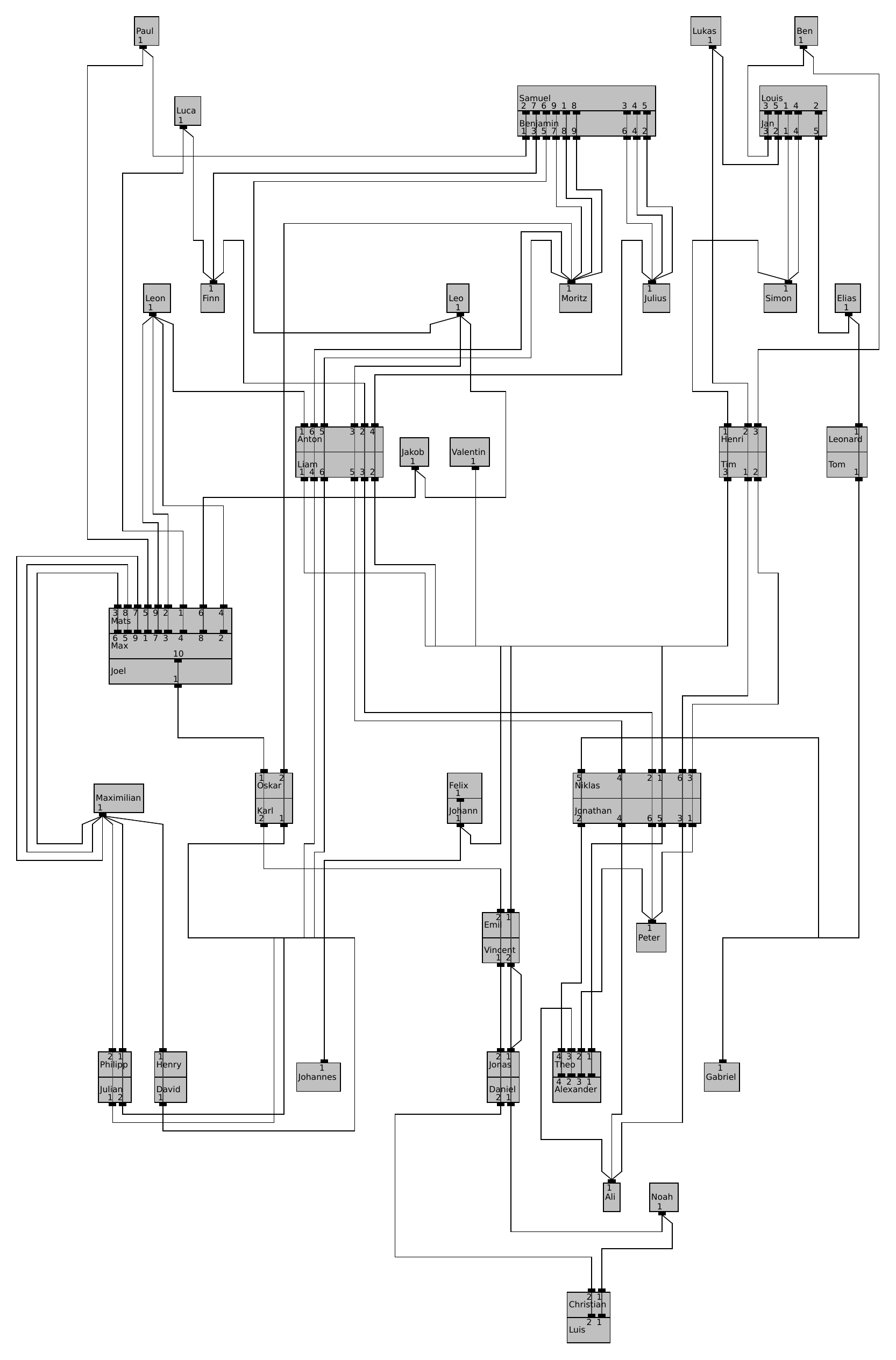}
	\caption{Original cable plan (anonymized) from the readable data set with 50 vertices and 52 crossings drawn by our algorithm using \fd, \ports and \relpos.}
	\label{fig:orig-readable-ports-50vtcs}
\end{figure}

\begin{figure}[p]
	\centering
	\includegraphics[height=.9\textheight]{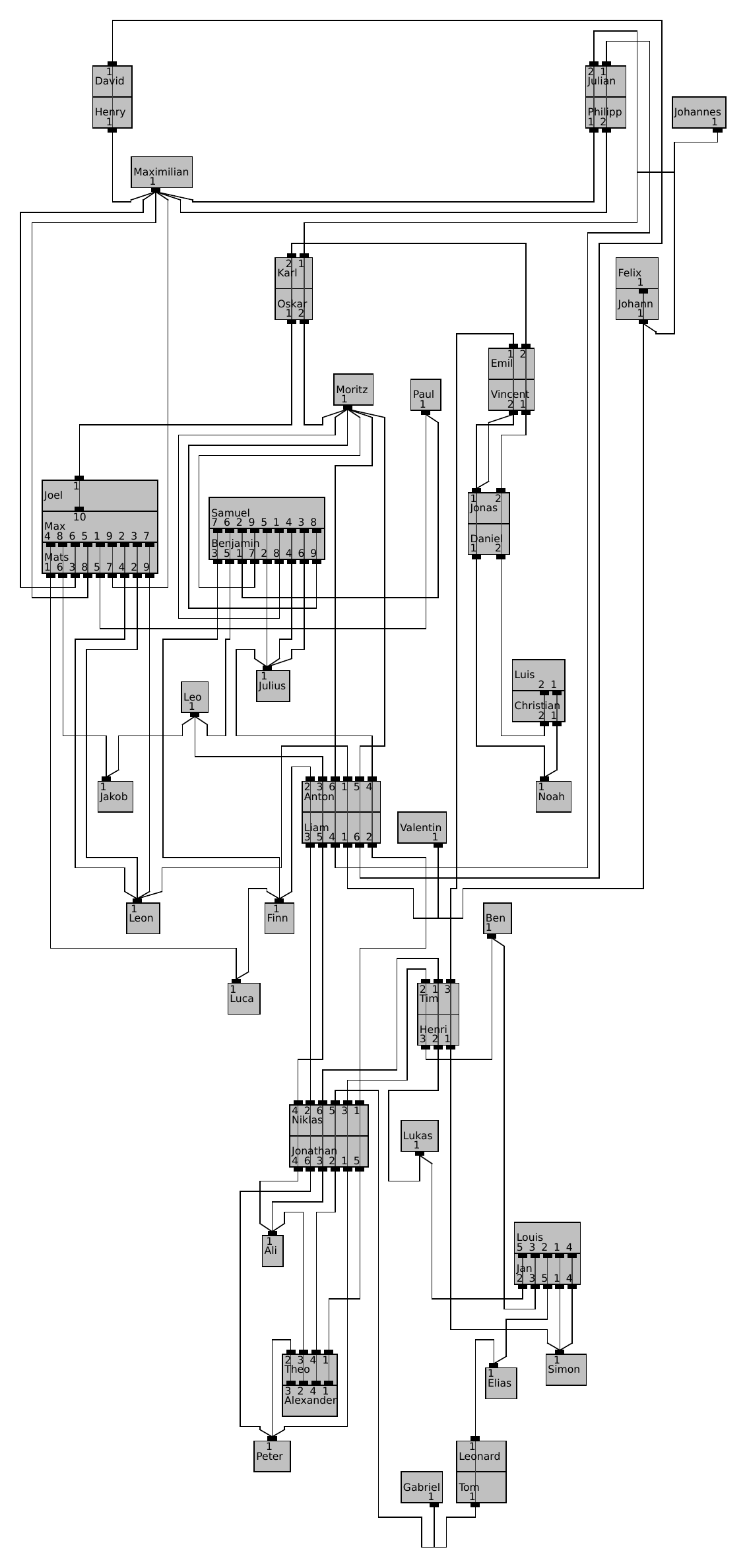}
	\caption{Original cable plan (anonymized) from the readable data set with 50 vertices and 71 crossings drawn using \kieler.}
	\label{fig:orig-readable-kieler-50vtcs}
\end{figure}

\begin{figure}[p]
	\centering
	\includegraphics[width=\textwidth]{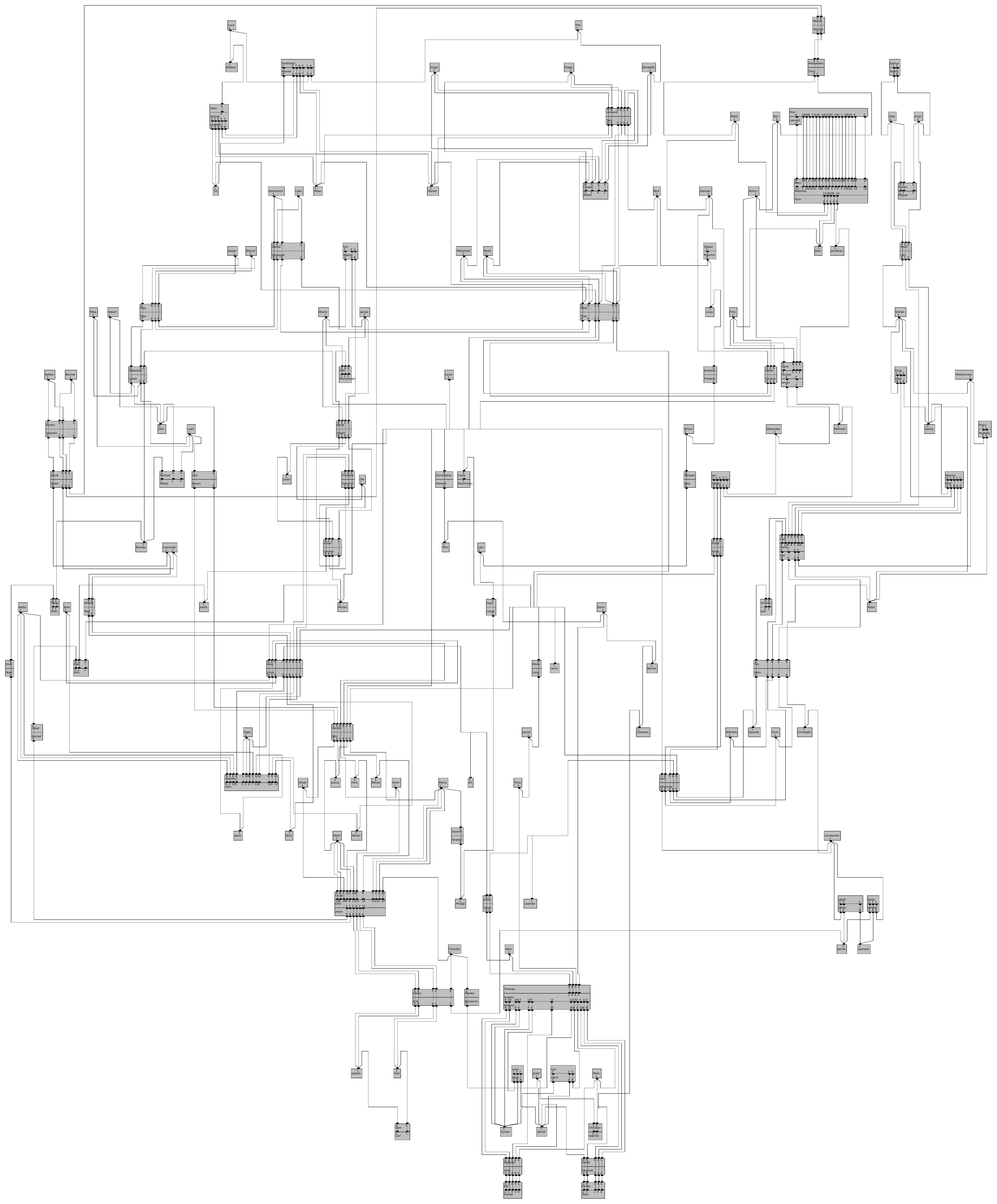}
	\caption{Original cable plan (anonymized) from the readable data set with 225 vertices and 391 crossings drawn by our algorithm using \fd, \ports and \relpos.}
	\label{fig:orig-readable-ports-246vtcs}
\end{figure}

\begin{figure}[p]
	\centering
	\includegraphics[height=.9\textheight]{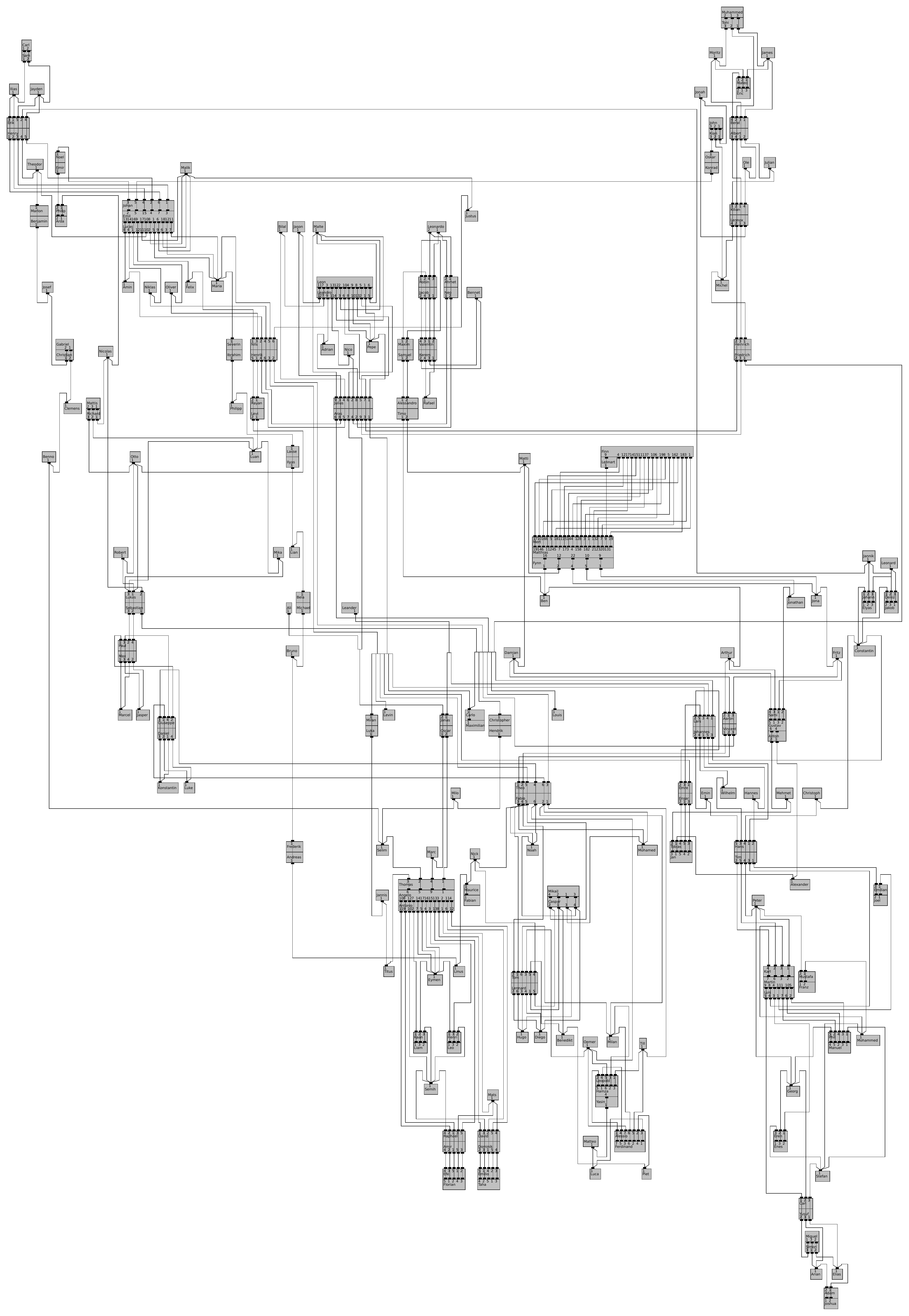}
	\caption{Original cable plan (anonymized) from the readable data set with 225 vertices and 562 crossings drawn using \kieler.}
	\label{fig:orig-readable-kieler-246vtcs}
\end{figure}

\begin{figure}[p]
	\centering
	\includegraphics[height=.9\textheight]{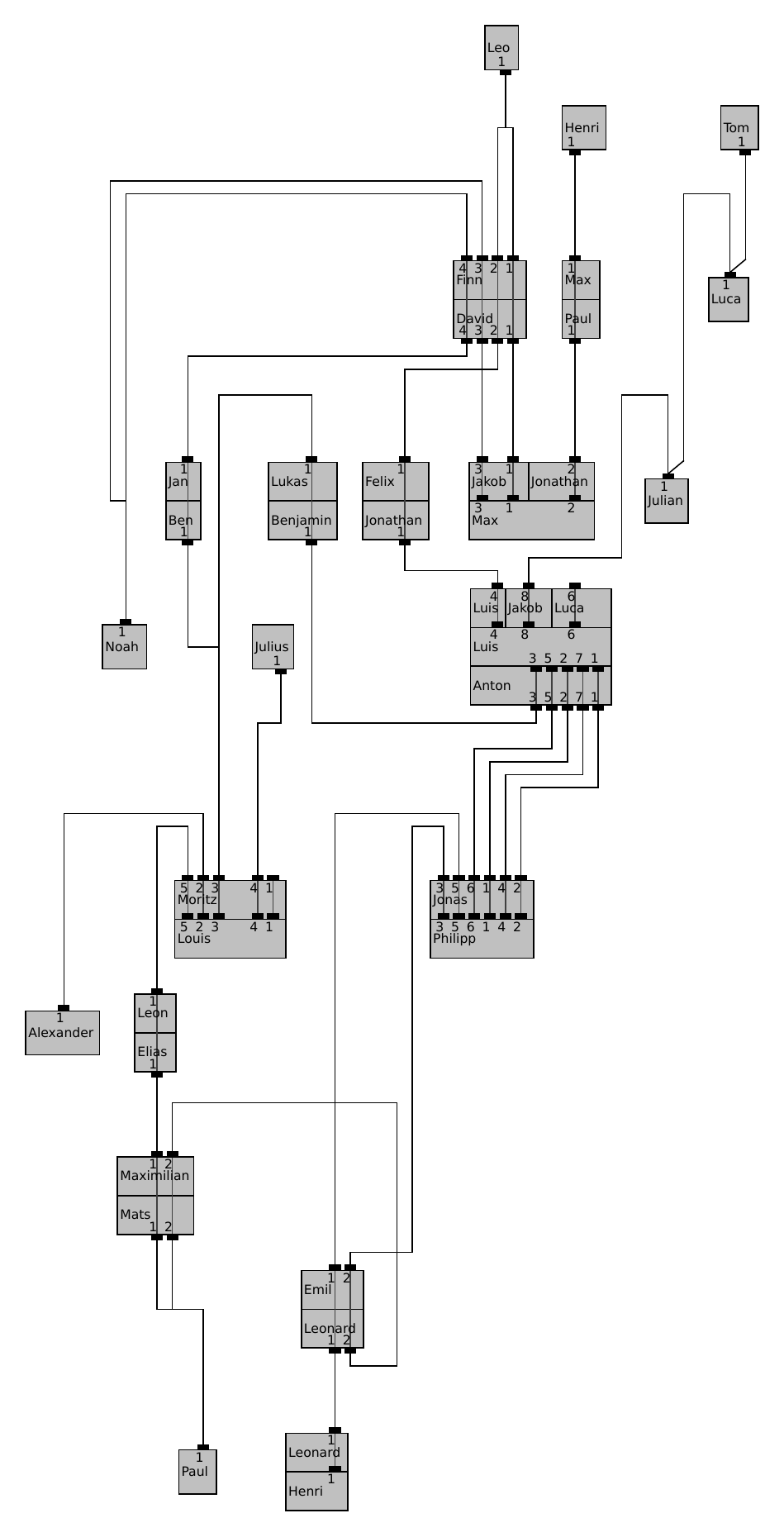}
	\caption{Pseudo cable plan generated from the readable data set with 39 vertices and three crossing drawn by our algorithm using \fd, \ports and \relpos.}
	\label{fig:pseudo-readable-ports-39vtcs}
\end{figure}

\begin{figure}[p]
	\centering
	\includegraphics[width=\textwidth]{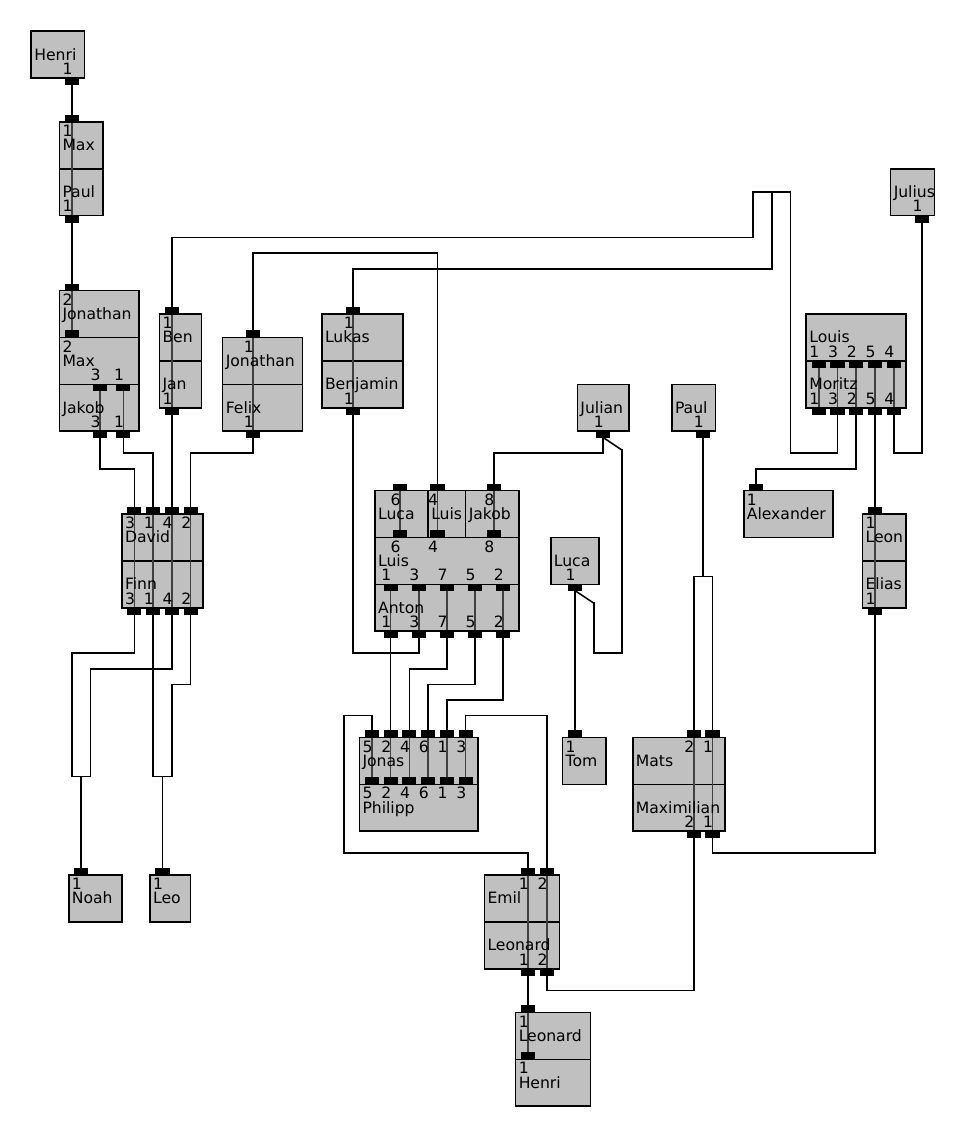}
	\caption{Pseudo cable plan generated from the readable data set with 39 vertices and three crossings drawn using \kieler.}
	\label{fig:pseudo-readable-kieler-39vtcs}
\end{figure}

\begin{figure}[p]
	\centering
	\includegraphics[height=.9\textheight]{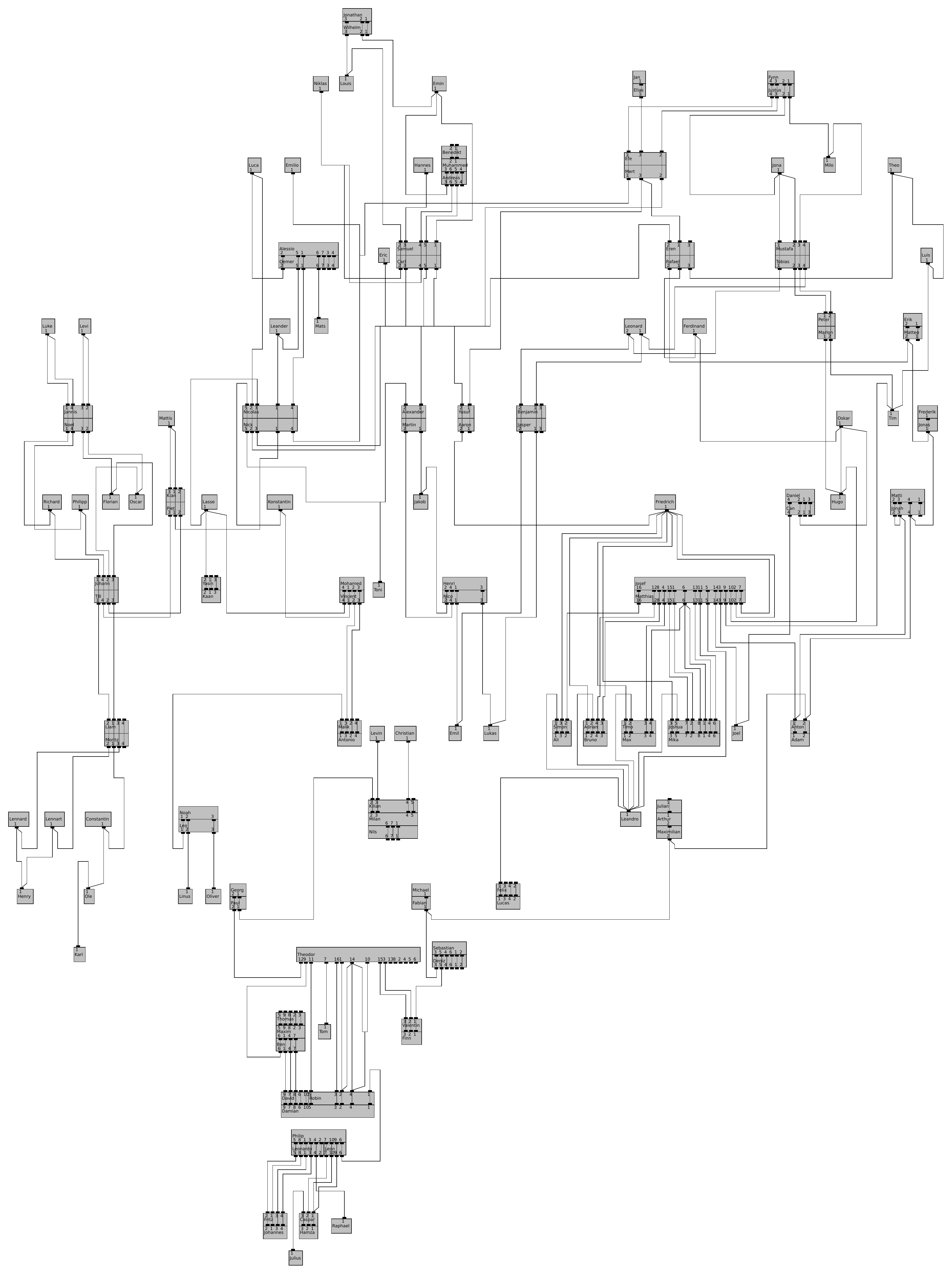}
	\caption{Pseudo cable plan generated from the readable data set with 144 vertices and 86 crossings drawn by our algorithm using \fd, \ports and \relpos.}
	\label{fig:pseudo-readable-ports-144vtcs}
\end{figure}

\begin{figure}[p]
	\centering
	\includegraphics[height=.9\textheight]{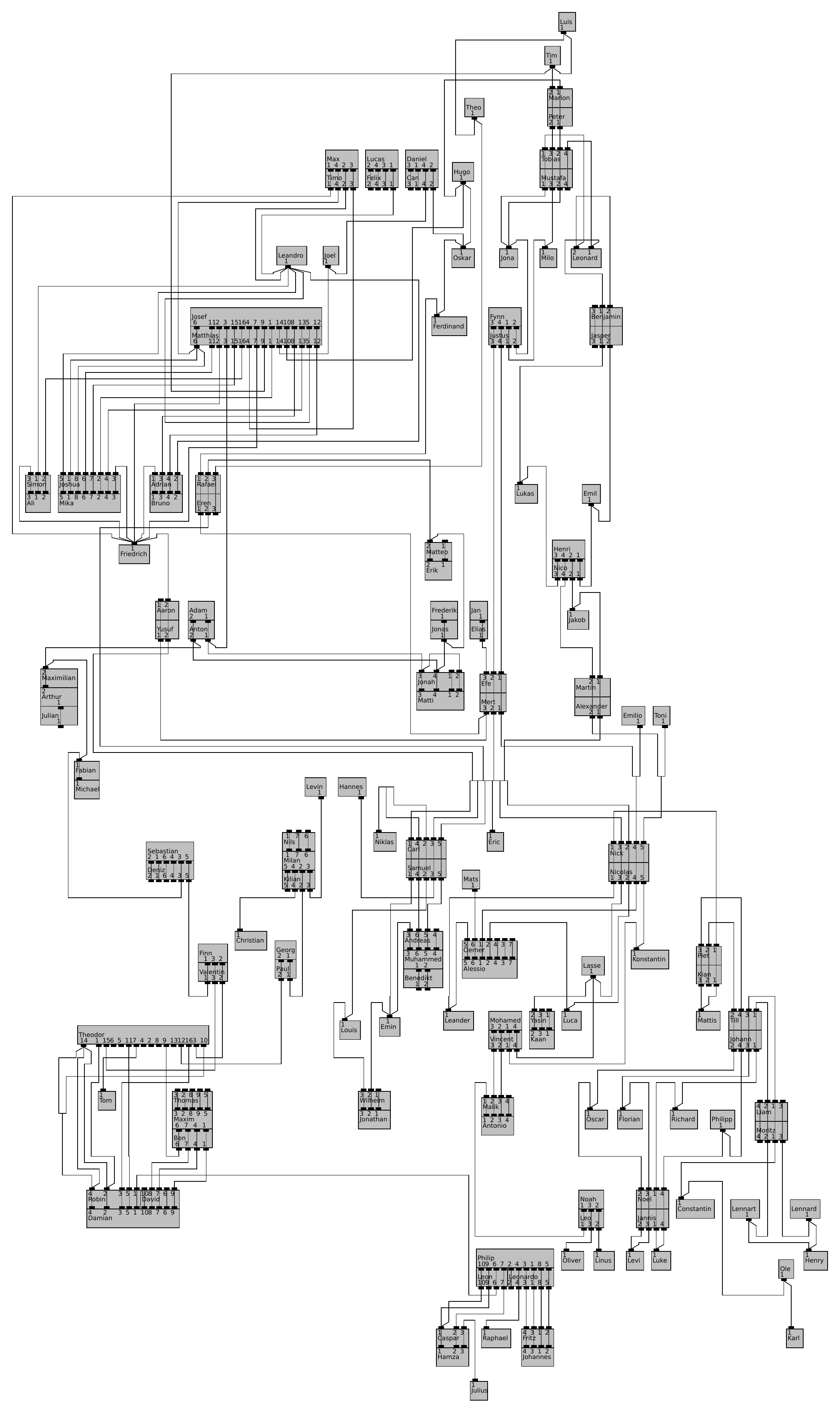}
	\caption{Pseudo cable plan generated from the readable data set with 144 vertices and 157 crossings drawn using \kieler.}
	\label{fig:pseudo-readable-kieler-144vtcs}
\end{figure}

\end{document}